\documentclass[numberedappendix]{emulateapj}
\usepackage{epsfig, natbib}

\tightenlines

\countdef\decade=200
\advance\decade by \year
\countdef\hours=201	
\advance\hours by \time
\divide\hours by 60
\countdef\mins=202
\advance\mins by \hours
\multiply\mins by 60
\multiply\hours by 100
\countdef\miltime=203
\advance\miltime by \hours
\advance\miltime by \time
\advance\miltime by -\mins

\newcommand{\msun}{M_{\odot}}

\begin{document}

\title{What Phase of the Interstellar Medium Correlates with the Star Formation Rate?} 


\shorttitle{ISM Phases and Star Formation}
\shortauthors{Krumholz et~al.}

\author{Mark R.\ Krumholz\altaffilmark{1}, Adam K.\ Leroy\altaffilmark{2}, and Christopher F.\ McKee\altaffilmark{3}}
\altaffiltext{1}{Department of Astronomy and Astrophysics, University of California, Santa Cruz, CA 95064, USA; krumholz@ucolick.org}
\altaffiltext{2}{Hubble Fellow; National Radio Astronomy Observatory, 520 Edgemont Rd., Charlottesvile, VA 22903, USA}
\altaffiltext{3}{Departments of Physics and Astronomy, University of California, Berkeley, CA 94720, USA} 

\begin{abstract}
Nearby spiral galaxies show an extremely tight correlation between tracers of molecular hydrogen (H$_2$) in the interstellar medium (ISM) and tracers of recent star formation, but it is unclear whether this correlation is fundamental or accidental. In the galaxies that have been surveyed to date, H$_2$ resides predominantly in gravitationally bound clouds cooled by carbon monoxide (CO) molecules, but in galaxies of low metal content the correlations between bound clouds, CO, and H$_2$ break down, and it is unclear if the star formation rate will then correlate with H$_2$ or with some other quantity. Here we show that star formation will continue to follow H$_2$ independent of metallicity. This is not because H$_2$ is directly important for cooling, but instead because the transition from predominantly atomic hydrogen (H~\textsc{i}) to H$_2$ occurs under the same conditions as a dramatic drop in gas temperature and Bonnor-Ebert mass that destabilizes clouds and initiates collapse. We use this model to compute how star formation rate will correlate with total gas mass, with mass of gas where the hydrogen is H$_2$, and with mass of gas where the carbon is CO in galaxies of varying metallicity, and show that preliminary observations match the trend we predict.
\end{abstract}


\section{Introduction}

Spiral galaxies near the Milky Way show strong correlation between the surface density of star formation and the surface density of molecular gas, but only a very weak correlation between star formation and the surface density of atomic gas \citep{kennicutt07a, bigiel08a, leroy08a}. However, the origin of this behavior is highly debated. In many models for the star formation rate \citep{tan00a, li05a, silk09a, dobbs09a}, the chemical state of the gas is assumed to be dynamically unimportant. In these models stars form wherever galactic-scale gravitational instabilities dictate that gas collects into gravitationally bound structures, without regard to whether those structures consist of atomic or molecular gas. The observed correlation between molecules and star formation occurs only because molecules form preferentially in the same bound structures where stars do. 

In other models \citep{robertson08b, gnedin09a, krumholz09b, pelupessy09a, gnedin10a, fu10a}, however, the underlying assumption is that star formation follows the chemical transition between atomic hydrogen (H~\textsc{i}) and molecular hydrogen (H$_2$) as the dominant gas phase, or the transition from ionized carbon (C~\textsc{ii}) to carbon monoxide (CO) as the primary coolant, independent of the global dynamics. Closely related to these are models in which the locations where stars can form is determined by the gas temperature, or by the interplay between gas temperature and gravitational stability \citep{elmegreen94c, schaye04a, ostriker10a}. It is not entirely clear how this is correlated with the chemical composition, although \citet{schaye04a} obtain the intriguing result, which anticipates to some extent the one we report here, that the formation of a cold phase is closely associated with the H~\textsc{i} to H$_2$ transition. However, he did not consider how this relates to CO, the molecule usually observed in place of H$_2$.

All of these models produce similar results for nearby spiral galaxies, where the majority of gravitationally-bound structures are also molecular and cooled by CO. However, whether star formation follows global dynamics, the H~\textsc{i} to H$_2$ transition, or the C~\textsc{ii} to CO transition makes a very large difference for how star formation behaves in regions with substantially sub-Solar metal content, since the chemical states of the hydrogen and the carbon are sensitive to gas metallicity in different ways \citep{van-dishoeck88a, krumholz09a, glover10b, wolfire10a}, while the prevalence of galactic-scale instabilities is very insensitive to metallicity. Moreover, any successful model must also be able to relate the star formation rate to observable quantities, which include H~\textsc{i} and CO, but not (generally) the total mass of H$_2$.

At present observations possess limited power to discriminate between the models, although the limited data available for nearby low-metallicity dwarf galaxies such as the Small Magellanic Cloud hold out the promise of being able to distinguish whether metallicity affects star formation. However, the differences between the models have profound implications for our understanding of star formation in the high redshift universe, where average metallicities are much lower. For example, the difficulty of forming molecules in metal-poor gas has been proposed as an explanation for the observed \citep{wolfe06a, wild07a} lack of star formation in damped Lyman $\alpha$ systems \citep{krumholz09e, gnedin10a}, but this explanation is only viable if transitions to molecular gas are necessary for star formation. 

Here we investigate whether the chemical state of the gas is relevant for star formation by studying how the Bonnor-Ebert mass and the chemical state vary in spherical interstellar clouds of varying volume and column densities. We show that, for such clouds, the H~\textsc{i} to H$_2$ transition is strongly associated with a dramatic drop in the temperature and Bonnor-Ebert mass within the cloud, across a very wide range of metallicities and environments. In Section~\ref{methods} we describe our computation method, in Section~\ref{results} we report our results, and in Section~\ref{observations} we discuss the implications of our results for observations. Finally, we summarize in Section \ref{summary}.

\section{Computation Method}
\label{methods}

To minimize clutter we summarize all the physical parameters that enter our model in Table \ref{tab:param}, and we discuss these choices in Appendix~\ref{app:paramdis}. We also show in that Appendix that our results are robust against changes in these parameters. In this section we limit ourselves to a discussion of our calculation method.

\begin{deluxetable*}{cccc}
\tabletypesize{\scriptsize}
\tablecaption{
\label{tab:param}
Fiducial parameter choices
}
\tablehead{
\colhead{Parameter} &
\colhead{Value} &
\colhead{Meaning} &
\colhead{Reference}
}
\startdata
$\kappa$ & $0.01(T/10\mbox{ K})^2 Z'$ cm$^2$ g$^{-1}$ & IR dust opacity & \citet{lesaffre05a} \\
$\alpha_{\rm gd}$ (H$_2$)  & 
$3.2\times 10^{-34}Z'$ erg cm$^3$ K$^{-3/2}$
& Dust-gas heat exchange rate & \citet{goldsmith01a} \\
$\alpha_{\rm gd}$ (H~\textsc{i})
& $1.0\times 10^{-33}Z'$ erg cm$^3$ K$^{-3/2}$ 
& Dust-gas heat exchange rate & Appendix \ref{app:paramdis}
\\
$q_{\rm CR}$ (H$_2$)
& 12.25 eV
& Energy / CR ionization & Appendix \ref{app:paramdis}
\\
$q_{\rm CR}$ (H~\textsc{i})
& 6.5 eV
& Energy / CR ionization & \citet{dalgarno72a}
\\ 
$\zeta$ & $2\times 10^{-17}Z'$ s$^{-1}$ & CR ionization rate & Appendix \ref{app:paramdis} \\
$T_{\rm rad}$ & 8 K & Ambient radiation temperature & Appendix \ref{app:paramdis} \\
$G_0'$ & 1 & UV radiation intensity & Appendix \ref{app:paramdis} \\
$X_{\rm CII}$, $X_{\rm CO}$ & $1.6\times 10^{-4} Z'$ & C~\textsc{ii} / CO abundance & \citet{sofia04a} \\
$\sigma_{d}$ & $1\times 10^{-21}Z'$ cm$^{-2}$ & UV dust opacity per H nucleus & Appendix \ref{app:paramdis} \\
$A_V/N_{\rm H}$ & $4.0\times 10^{-22}Z'$ mag cm$^2$ & Visual extinction per H column & Appendix \ref{app:paramdis} \\
OPR & 0.25 & Ratio of ortho- to para-H$_2$ & \citet{neufeld06a} \\
$\sigma$ & 2 km s$^{-1}$ & Cloud velocity dispersion & \citet{krumholz07g}
\enddata
\tablecomments{References to Appendix \ref{app:paramdis} mean that parameter choices are discussed there.}
\end{deluxetable*}

\subsection{Physical Model}

Consider a spherical cloud of mean volume density $n_{\rm H}$ H nuclei cm$^{-3}$ and mean column density $N_{\rm H}$ H nuclei cm$^{-2}$, mixed with dust with an absorption cross section per H nucleus $\sigma_{d}$ to photons near 1000 \AA. The mean absorption optical depth to UV photons is $\tau = N_{\rm H} \sigma_d$, and the corresponding mean visual extinction at optical wavelengths is $A_V$. The outer parts of the cloud are exposed to the ultraviolet interstellar radiation field (ISRF) of the galaxy, and this keeps the hydrogen and carbon in the outer parts of the cloud predominantly in the form of H~\textsc{i} and C~\textsc{ii}. If the column density is sufficiently large, dust and the small population of hydrogen molecules in the predominantly atomic layer will absorb the ISRF, allowing a transition from H~\textsc{i} to H$_2$ and C~\textsc{ii} to CO as the main repositories of hydrogen and carbon. We assume that the gas is neutral, and that there are very few ionizing photons present.

In a real interstellar cloud, even if the density were uniform, the temperature and chemical composition would not be. Surface layers exposed to direct UV radiation would be warmer than the shielded interior, and the hydrogen there would be primarily H~\textsc{i} rather than H$_2$. Thus there is no single cloud temperature or chemical composition we can compute, and determining the full temperature and chemical profile even when the density profile is given in advance requires solving a photodissociation region model in which one both simultaneously determines temperature and the chemical state, including line radiative transfer throughout the cloud. While there are many examples of such models in the literature, they are too computationally expensive to allow a wide search of parameter space, and they still rely on arbitrarily chosen density distributions. Instead, our goal is to estimate a single characteristic temperature and for a cloud, coupled to a simple description of its chemistry. To this end we will describe clouds' chemical composition with single numbers giving the mass fraction in a given chemical phase, and we will approximate the cloud interior as a uniform region of volume density $n_{\rm H}$ and mean column density $N_{\rm H}$, consisting of gas at temperature $T_g$ uniformly mixed with dust of temperature $T_d$. For the purposes of computing the temperature, we will take the chemical composition to be uniform, with the hydrogen all as H~\textsc{i} or H$_2$, and the carbon all as C~\textsc{ii} or CO, despite the fact that every H$_2$ cloud has an outer H~\textsc{i} shell, and every CO cloud has an outer C~\textsc{ii} shell.

\subsection{Chemical State}

To compute the H~\textsc{i} to H$_2$ transition in our model cloud, we rely on the models of \citet{krumholz08c, krumholz09a}, and \citet{mckee10a} (collectively the KMT model hereafter). In these models the fraction of the hydrogen mass in the H$_2$-dominated region is given by
\begin{equation}
\label{eq:fh2}
f_{\rm H_2} \approx 1 - \left(\frac{3}{4}\right) \frac{s}{1+0.25s}
\end{equation}
where
\begin{eqnarray}
s & = & \frac{\ln(1+0.6\chi+0.01\chi^2)}{0.6\tau} \\
\chi & = & 71 \frac{G_0'}{n_{\rm H}/\mbox{cm}^{-3}}
\end{eqnarray}
and $G_0'$ is the strength of the ISRF normalized to its value in the Solar vicinity (which corresponds to a free-space H$_2$ dissociation rate of $5.4\times 10^{-11}$ s$^{-1}$). Note that, unlike some of the other approximate expressions derived by KMT, this one applies independent of whether the gas is warm or cold neutral medium. It should also be noted that this model assumes chemical equilibrium, which is not necessarily the case for H$_2$ \citep{pelupessy09a, mac-low10a}. However, \citet{krumholz11a} perform a detailed comparison between the equilibrium approximation and a set of galaxy simulations including full time-dependent chemistry and radiation. They find that the approximation is very accurate at metallicities of $Z'\ga 0.01$, where $Z'$ is the metallicity normalized to the Milky Way value, and we can therefore use it safely. We note that equation (\ref{eq:fh2}) also agrees extremely well with observations both of local galaxies \citep{krumholz09a} and of damped Lyman $\alpha$ systems at high redshift \citep{krumholz09c}. We therefore conclude that, on the galactic scales relevant to this work (as opposed to the isolated periodic boxes considered by \citealt{mac-low10a}) this approximation is valid.

The C~\textsc{ii} to CO transition is predominantly governed by dust extinction, and thus is sensitive primarily to the cloud optical depth. At high optical depths wherever the hydrogen is H$_2$ the carbon is also CO, while at low optical depths there can be significant regions of H$_2$ where the carbon is primarily C~\textsc{ii}, because the H$_2$ self-shields while the CO cannot \citep{glover10b, wolfire10a}. Recent semi-analytic work indicates that the ratio of mass where the carbon is CO to total cloud mass is \citep{wolfire10a}
\begin{equation}
f_{\rm CO} = f_{\rm H_2} e^{-4.0 \left(0.53 - 0.045\ln\frac{G_0'}{n_{\rm H}/{\rm cm}^{-3}}-0.097\ln Z'\right)/A_V}.
\label{eq:fco}
\end{equation}
Alternately, an empirical fit to numerical simulations for the same quantity \citep{glover10b} gives very similar results.

\subsection{Thermal State}

The temperatures of gas and dust are set by the condition of thermal equilibrium, following a method that combines elements from \citet{goldsmith01a} and \citet{lesaffre05a}. We consider heating of the gas by the grain photoelectric effect at a rate per H nucleus $\Gamma_{\rm pe}$, heating by cosmic rays at a rate $\Gamma_{\rm cr}$, and cooling via atomic and molecular line emission at a rate $\Lambda_{\rm line}$. Dust is heated via an external radiation field at a rate $\Gamma_{\rm dust}$, and cools via thermal emission at a rate $\Lambda_{\rm dust}$. Finally, energy flows from the dust to the gas at a rate $\Psi_{\rm gd}$, which may be negative if the gas is hotter than the dust. The condition for thermal balance is therefore that the temperatures $T_g$ and $T_d$ simultaneously satisfy the equations
\begin{eqnarray}
\label{eq:therm1}
\Gamma_{\rm pe} + \Gamma_{\rm cr} - \Lambda_{\rm line} + \Psi_{\rm gd} & = & 0 \\
\label{eq:therm2}
\Gamma_{\rm dust} - \Lambda_{\rm dust} - \Psi_{\rm gd} & = & 0.
\end{eqnarray}
Note that we do not include photoionization heating. While this is important at low column densities \citep[e.g.][]{schaye04a}, the lowest column density clouds we will consider in this work have optical depths of $\sim 10^3$ to ionizing photons, and thus photoionization heating is unimportant for them except very close to their surfaces.

To solve this equation we must compute the temperature dependence of each heating and cooling term. We assume that the clouds in question are optically thin to far-infrared radiation, so we need not consider trapping of the dust radiation field. Thus the dust cooling rate is
\begin{equation}
\Lambda_{\rm dust} = \kappa(T_d) \mu_{\rm H} c a T_d^4,
\end{equation}
where $c$ is the speed of light, $a$ is the radiation constant, $\kappa(T_d)$ is the temperature-dependent dust specific opacity and $\mu_{\rm H}$ is the mean mass per H nucleus. Dust heating is more complex, as mentioned above, since the external interstellar ultraviolet field dominates at cloud edges and the local re-radiated infrared field dominates at cloud centers. Since we are mostly interested in what happens when clouds are shielded by at least a few magnitudes of extinction in the ultraviolet, we choose to adopt the IR-dominated case. In practice this choice makes very little difference, since the dust gas coupling is a very minor effect at the densities where we focus our attention. We characterize the IR radiation field by an effective temperature $T_{\rm rad}$, so
\begin{equation}
\Gamma_{\rm dust} = \kappa(T_{\rm rad}) \mu_{\rm H} c a T_{\rm rad}^4.
\end{equation}
Finally, for the dust-gas energy exchange term we adopt the approximate exchange rate \citep{goldsmith01a}
\begin{equation}
\Psi_{\rm gd} = \alpha_{\rm gd} n_{\rm H} T_g^{1/2} (T_d - T_g),
\end{equation}
where $\alpha_{\rm gd}$ is a coupling constant.

For the gas, the cosmic ray heating rate is
\begin{equation}
\Gamma_{\rm cr} = \zeta q_{\rm CR}\mbox{ s}^{-1}
\end{equation}
where $\zeta$ is the cosmic ray primary ionization rate and $q_{\rm CR}$ is the thermal energy increase per primary cosmic ray ionization. For grain photoelectric heating, we must choose a suitable mean heating rate, since extinction through the cloud causes the heating rate to drop substantially as we move toward the interior. For simplicity, and since we are concerned more with cloud interiors than surfaces, we consider the heating rate to be attenuated by half the mean extinction of the cloud. With this approximation, we have
\begin{equation}
\Gamma_{\rm pe} = 4.0\times 10^{-26} G_0' e^{-N_{\rm H} \sigma_{d}/2}\mbox{ erg s}^{-1},
\end{equation}
where $\sigma_{d}$ is the dust cross section per H nucleus to the UV photons that dominate grain photoelectric heating.

The final term in the thermal balance equations (\ref{eq:therm1}) and (\ref{eq:therm2}) is $\Lambda_{\rm line}$, the line cooling rate. In each case we consider only a single coolant species: C~\textsc{ii} or CO. This is a reasonable approximation because, at the low temperatures where star formation occurs, one of these two will dominate. To compute the cooling rate we must compute the populations of the various levels, and we do so following the method of \citet{krumholz07g}. Let $f_i$ be the fraction of a given species, C~\textsc{ii} or CO, in the $i$th state, and let $E_i$ be the energy of that state. Transitions between states $i$ and $j$ occur due to spontaneous emission, at a rate given by the Einstein coefficient $A_{ij}$ (which is zero if $E_i \leq E_j$), and due to collisions with with rate coefficient $\gamma_{ij,\rm S}$, where $S$ is the species of the collision partner. Note that the collision rate coefficients depend on the gas temperature. We do not consider stimulated emission or absorption due to an external radiation field, since neither is significant for the cooling lines of C~\textsc{ii} or CO.

In the escape probability formalism we approximate that each spontaneously emitted photon produced by an atom transiting from level $i$ to level $j$ has a probability $\beta_{ij}$ of escaping; photons that do not escape are re-absorbed on the spot within the cloud, yielding no net change in the level populations. With these definitions and approximations, in statistical equilibrium the level populations are given implicitly by
\begin{eqnarray}
f_i \sum_j \left(\beta_{ij} A_{ij} + n_{\rm HI} \gamma_{ij,\rm HI} + n_{\rm He} \gamma_{ij,\rm He} \right.
\nonumber \\
\left. {} + n_{\rm p- H_2}\gamma_{ij,\rm p-H_2}  + n_{\rm o-H_2} \gamma_{ij,\rm o-H_2}\right) =
\nonumber \\
\sum_i \left(\beta_{ij} A_{ij} + n_{\rm HI} \gamma_{ij,\rm HI} + n_{\rm He} \gamma_{ij,\rm He} 
\right.
\nonumber\\
\left. {}
+ n_{\rm p-H_2}\gamma_{ij,\rm p-H_2} + n_{\rm o-H_2} \gamma_{ij,\rm o-H_2}\right) f_i,
 \label{eq:levpop}
\end{eqnarray}
where the sums run over all quantum states, and $n_{\rm HI}$, $n_{\rm p-H_2}$, $n_{\rm o-H_2}$, and $n_{\rm He}$ are the number densities of H~\textsc{i}, para-H$_2$, ortho-H$_2$, and He, respectively. The left hand side of this equation represents the rate of transitions out of state $i$ to all other states $j$, while the right hand side represents the rate of transitions into state $i$ from all other states $j$. The escape probabilities themselves depend on the level population. If we let $X$ be the abundance of the species in question relative to H nuclei, and $\sigma$ be the one-dimensional gas velocity dispersion, then we have \citep{krumholz07g}
\begin{equation}
\beta_{ij} \approx \frac{1}{1+0.5\tau_{ij}},
\end{equation}
where
\begin{equation}
\label{eq:tauij}
\tau_{ij} = \frac{g_i}{g_j} \frac{3A_{ij} \lambda_{ij}^3}{16(2\pi)^{3/2} \sigma} X N_{\rm H} f_j \left(1 - \frac{f_i g_j}{f_j g_i}\right),
\end{equation}
$g_i$ and $g_j$ are the statistical weights of levels $i$ and $j$, and $\lambda_{ij}$ is the wavelength of a photon associated with the transition from level $i$ to level $j$.

Equations (\ref{eq:levpop}) -- (\ref{eq:tauij}) constitute a complete system of algebraic equations for the level populations $f_i$. Given the solution to these equations, the line cooling rate per H nucleus is then given by
\begin{equation}
\Lambda_{\rm line} = \sum_{i,j} \beta_{ij} A_{ij} f_i h \nu_{ij} / n_{\rm H},
\end{equation}
where $\nu_{ij}$ is the frequency of a photon emitted in a transition from state $i$ to state $j$.

We have now written down a complete set of equations to determine the gas and radiation temperatures. To solve them we use a double-iteration method. We first select trial values of $T_d$ and $T_g$, and then we compute all the heating and cooling rates $\Gamma$, $\Lambda$, and $\Psi$. In order to compute $\Lambda_{\rm line}$, we must solve for the equilibrium level populations $f_i$ by solving equations (\ref{eq:levpop}) -- (\ref{eq:tauij}). We do so by fixing $T_g$ (and thus all the collision rate coefficients) and applying a Broyden's method \citep{press92a}. Once we have determined $\Lambda_{\rm line}$, we check if the thermal balance equations (\ref{eq:therm1}) and (\ref{eq:therm2}) are satisfied to within a specified tolerance. If not, we iteratively update $T_d$ and $T_g$ using Newton's method to update $T_d$ and $T_g$ until equations (\ref{eq:therm1}) and (\ref{eq:therm2}) are satisfied to the required tolerance.

\section{Results}
\label{results}

\begin{figure}
\epsscale{1.2}
\plotone{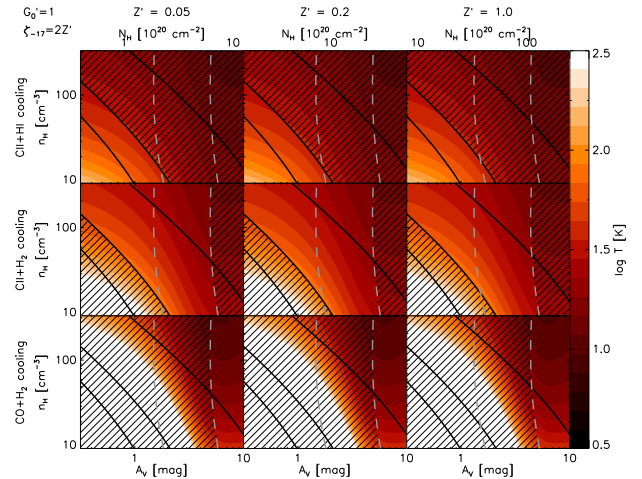}
\epsscale{1.0}
\caption{
\label{fig:tempgrid}
Color indicates gas temperature as a function of density $n_{\rm H}$ and visual extinction $A_V$ for metallicities of $Z'=0.05$, $0.2$, and $1.0$ (left to right columns; prime indicates relative to the Solar neighborhood value) and cooling by C~\textsc{ii} plus H~\textsc{i} (top row), C~\textsc{ii} plus H$_2$ (middle row), and CO plus H$_2$ (bottom row) using our fiducial parameters $G_0' = 1$, $\zeta_{-17} = \zeta / 10^{-17}\mbox{ s}^{-1} = 2Z'$ (upper left corner). Color contours run from $\log (T/\mbox{K}) = 0.5$ to $\log (T/\mbox{K})=2.5$ in steps of 0.1, with white representing $\log (T/\mbox{K}) \geq 2.5$. The black solid lines are contours of $f_{\rm H_2}=0.1$, $0.5$, and $0.9$, from left to right. The gray dashed lines are contours of $f_{\rm CO} = 0.1$ and $0.5$ from left to right; the $f_{\rm CO}=0.9$ contour lies off the plot to the right. Finally, the hatching pattern marks regions where the temperature and chemistry calculations are not fully consistent (e.g.\ because the chemistry calculation indicates that a cloud of that $A_V$ and $n$ should be dominated by H$_2$, but the temperature calculation assumes the hydrogen is H~\textsc{i}).
}
\end{figure}

We use the procedure outlined in Section \ref{methods} to compute the chemical state and temperature for a grid of clouds of varying $n_{\rm H}$ and $A_V$. Figure \ref{fig:tempgrid} shows the equilibrium temperature as a function of $n_{\rm H}$ and $A_V$, overlaid with contours of H$_2$ and CO fraction, computed for the three possible chemical compositions and gas metallicities of $Z' = 0.05$, $0.2$, and $1.0$. Recall that the chemical and thermal computations are decoupled, so the temperatures for C~\textsc{ii} plus H~\textsc{i} cooling should only be considered reliable to the left of $f_{\rm CO} = 0.5$ and $f_{\rm H_2} = 0.5$ contours, indicating the C~\textsc{ii} to CO and H~\textsc{i} to H$_2$ transitions, respectively. Similarly, the CO plus H$_2$ temperatures are only reliable to the right of both of these curves, while the C~\textsc{ii} plus H$_2$ temperatures are reliable in the region between them. 

Regardless of these limits on the regions of applicability, the striking result from these plots is that, {\it independent} of metallicity or chemical composition, there is a dramatic drop in temperature from hundreds of K to $\sim 10$ K as one moves from lower left (low density, low $A_V$) to upper right (high density, high $A_V$), and that the contours of constant temperature align remarkably well with contours of constant $f_{\rm H_2}$. This result is not surprising, because temperature and H$_2$ fraction depend on density and extinction in very similar ways. Both the heating and H$_2$ dissociation rate are proportional to the the exponential of minus the visual extinction, and both the cooling and H$_2$ formation rates are proportional to the square of the volume density. In contrast, contours of constant CO align much less well with temperature contours. The CO fraction depends primarily on $A_V$, with only a very weak density dependence. Furthermore, at $A_V$ high enough for the CO fraction to reach 50\%, the grain photoelectric effect has been shut off so thoroughly that the temperature is insensitive to further increases in $A_V$.

\begin{figure}
\epsscale{1.2}
\plotone{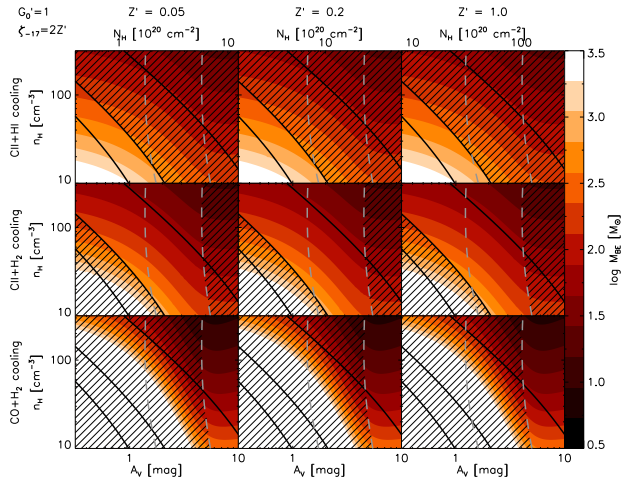}
\epsscale{1.0}
\caption{
\label{fig:mjgrid}
Same as Figure \ref{fig:tempgrid}, but with color indicating Bonnor-Ebert mass instead of temperature. Color contours run from $\log(M_{\rm BE}/\msun) = 0.5$ to $\log(M_{\rm BE}/\msun) = 3.5$ in steps of 0.25, with white representing $\log(M_{\rm BE}/\msun)\geq 3.5$.
}
\end{figure}

\begin{figure}
\epsscale{1.2}
\plotone{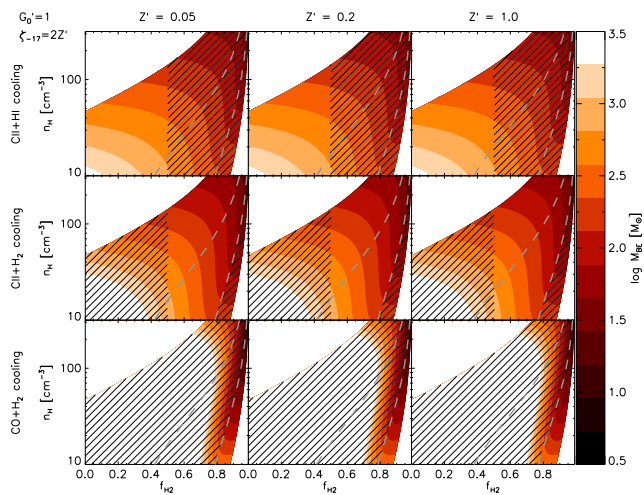}
\epsscale{1.0}
\caption{
\label{fig:mjh2}
Contours are the same as in Fig.\ 2, but the $x$ axis shows H$_2$ fraction rather than $A_V$. Note that not all of the grid is colored, because our grid in $n_{\rm H}$ and $A_V$ does not cover the full range in $n_{\rm H_2}$ and $f_{\rm H_2}$ shown.
}
\end{figure}

Figures \ref{fig:mjgrid} and \ref{fig:mjh2} show how the Bonnor-Ebert mass, the largest mass that can be supported against collapse by thermal pressure, changes with $A_V$ and with H$_2$ fraction as a result of this temperature change. This mass is
\begin{equation}
M_{\rm BE} = 1.18 \frac{c_s^3}{\sqrt{G^3 n \mu_{\rm H}}} = 1.18 \left[\frac{(k_B T/ \mu)^3}{G^3 n \mu_{\rm H}}\right]^{1/2},
\end{equation}
where $c_s$ is the isothermal sound speed, $T$ is the temperature, and $\mu_{\rm H}$ and $\mu$ are the mean mass per H nucleus and the mean particle mass, respectively. For Milky Way helium abundance, the former is $\mu_{\rm H}=2.3\times 10^{-24}$ g regardless of chemical composition, while the latter is $2.1\times 10^{-24}$ g for H~\textsc{i} and $3.8\times 10^{-24}$ g for H$_2$. As the plot shows, there is a radical drop in the Bonnor-Ebert mass from thousands of $\msun$ to a few $\msun$ as one moves from low density and extinction to high density and extinction, and, as with the temperature, there is a very strong correlation between Bonnor-Ebert mass and H$_2$ fraction, while there is relatively little correlation with CO fraction. Figure 3 clearly shows that small Bonnor-Ebert masses are found exclusively in clouds with high H$_2$ fractions.

Our results imply that, in gas whose conditions are such that the hydrogen is mostly H~\textsc{i}, structures with masses of $\sim 1000$ $\msun$ or less will  be stabilized against collapse by thermal pressure. As a result, we conclude that star formation in such environments is extremely unlikely, in part because turbulence is extremely unlikely to generate fragments of such high masses that could then collapse \citep{padoan02a}. Conversely, in gas where the hydrogen is mostly H$_2$, the mass that can be stabilized by thermal pressure is $2-3$ orders of magnitude smaller, turbulence generates numerous fragments capable of collapsing, and star formation is far more likely. It is important to note that the drop in Bonnor-Ebert mass and loss of stability is not caused by the H~\textsc{i} to H$_2$ transition, it is simply very well correlated with it because the temperature and the H$_2$ fraction are determined by very similar combinations of extinction and density. We therefore conclude that star formation should correlate well with H$_2$, simply because H$_2$ is a good tracer of regions where thermal pressures are low enough to allow star formation.

\section{Observational Consequences}
\label{observations}

\subsection{Predictions for Observable Quantities}

We can use our result that star formation correlates with H$_2$ to predict an approximate correlation between star formation rates and the masses of all gas, H$_2$ gas, and CO gas as a function of galaxy metallicity and surface density. Consider a portion of a galaxy with a mean surface density $\Sigma_g$ (averaged over the $\sim$kpc scales accessible to current observations). Within this region a fraction $f_{\rm H_2}$ of the gas, given by equation (\ref{eq:fh2}), is in the form of star-forming H$_2$ clouds, and within these a smaller fraction $f_{\rm CO}$ of the gas, given by equation (\ref{eq:fco}), has most of its carbon in CO molecules.

We evaluate $f_{\rm H_2}$ following the methods outlined in \citet{krumholz09a, krumholz09b}. First, we must adopt a clumping factor $c$ to scale from the surface densities of individual atomic molecular complexes on $\sim 100$ pc scales to the $\sim$kpc scales accessible to current observations. This is necessary because the mean surface density averaged over a kpc-scale region that we observe is lower than the surface densities of the $\sim 100$ pc-sized giant molecular clouds, but it is the latter rather than the former that determines the atomic to molecular ratio in these clouds. Based on a combination of theoretical arguments and fits to observation, \citet{krumholz09a} adopt $c=5$, and we do so here, so $\tau = 5 \Sigma_g \sigma_d / \mu_{\rm H}$. Second, we adopt a characteristic value of $G_0'/n = 0.044 (1 + 3.1 Z'^{0.365})/4.1$ expected for a two-phase atomic medium \citep{krumholz09a, krumholz09b}. (For a given value of $G_0'$, this corresponds to assuming that clouds on Figures \ref{fig:tempgrid} -- \ref{fig:mjh2} form a horizontal sequence at a fixed $n$ value that depends on $G_0'$ and $Z'$.) To evaluate $f_{\rm CO}$ and the star formation rate, we note that star-forming H$_2$ clouds appear to develop column densities $N_{\rm H}\sim 7.5\times 10^{21}$ cm$^{-2}$ ($\sim 85$ $\msun$ pc$^{-2}$) independent of galactic environment \citep{heyer09a, bolatto08a, krumholz06d}. Since the CO-dominated regions are the inner parts of these clouds, this implies that the column density and $A_V$ that enter the calculation of the CO fraction should not be the mean galactic surface density or extinction, but instead the fixed $A_V$ of star-forming H$_2$ clouds. We adopt this value for $A_V$ in equation (\ref{eq:fco}). 

The corresponding star formation rate is expected on theoretical grounds to be approximately \citep{krumholz09b}
\begin{eqnarray}
\dot{\Sigma}_* & = & \frac{f_{\rm H_2} \Sigma_g}{2.6\mbox{ Gyr}} \times
\nonumber \\
& & 
\left\{
\begin{array}{ll}
(\Sigma_g/85\,\msun\,{\rm pc}^{-2})^{-0.33}, & \Sigma_g < 85 \,\msun\,{\rm pc}^{-2} \\
(\Sigma_g/85\,\msun\,{\rm pc}^{-2})^{0.33}, & \Sigma_g > 85 \,\msun\,{\rm pc}^{-2} \\
\end{array}
\right..
\end{eqnarray}
In reality, based on the argument we have just made, we should compute the star formation rate not based directly on the H$_2$ fraction but instead based on the mass of cold gas. However, we have already seen that the H$_2$ and cold gas fractions are very similar, and equation (\ref{eq:fh2}) provides a convenient analytic approximation. We therefore use it to estimate both $f_{\rm H_2}$ and the star-forming gas fraction. We have therefore computed, for a portion of a galaxy of total gas surface density $\Sigma_g$ and metallicity $Z'$, the expected surface densities of H$_2$, CO, and star formation.

\begin{figure}
\epsscale{1.2}
\plotone{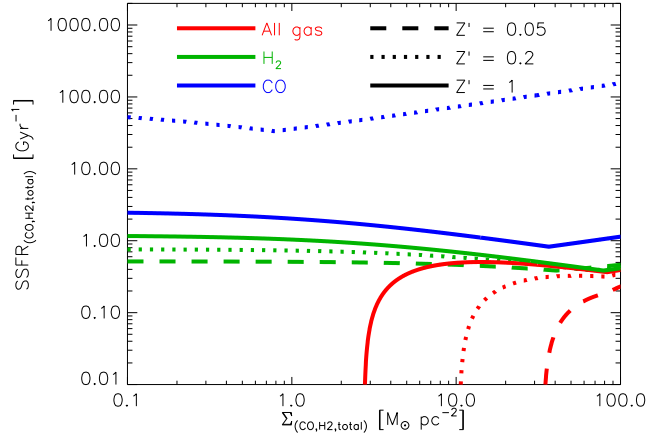}
\epsscale{1.0}
\caption{
\label{fig:sflaw1}
Predicted gaseous specific star formation rate ${\rm SSFR}_{(\rm CO,H_2,total)} = \dot{\Sigma}_*/\Sigma_{(\rm CO,H_2,total)} = \dot{M}_*/M_{(\rm CO,H_2,total)}$ as a function of gas surface density at fixed metallicity. The red lines indicate the specific star formation rate considering the total gas mass, green lines indicate the specific star formation rate for H$_2$, while blue lines indicate the specific star formation rate for gas where the carbon is CO rather than C~\textsc{ii}. Dashed lines correspond to a metallicity $Z'=0.05$, dotted lines to $Z'=0.2$, and solid lines to $Z'=1$. Note that the dashed blue line (${\rm SSFR}_{\rm CO}$ for $Z'=0.05$) is above the range shown in the plot. 
}
\end{figure}

Figure \ref{fig:sflaw1} shows our predicted correlation between specific star formation rate with respect to the mass of each gas constituent, $\mbox{SSFR}_{\rm(CO,H_2,total)} = \dot{\Sigma}_* / \Sigma_{\rm(CO,H_2,total)}$, and the surface density of that component at a range of metallicities. We see that at high surface densities and high metallicities the specific star formation rates for total gas, H$_2$, and CO are essentially identical, consistent with observations. At lower surface densities or metallicities, though, the star formation rate per unit total gas falls. Conversely, at lower metallicity the specific star formation with respect to CO rises, reflecting the fact that, at lower metallicity, the mass of gas where the chemical makeup is H$_2$ plus C~\textsc{ii} rises as a fraction of the total H$_2$ mass. Since star formation follows H$_2$, $\mbox{SSFR}_{\rm CO}$ rises as a result. We note that \citet{pelupessy09a} reached a similar conclusion about $\mbox{SSFR}_{\rm CO}$ based on simulations that used a star formation recipe that assumed roughly constant $\mbox{SSFR}_{\rm H_2}$.

\subsection{Comparison to Observations}

The most striking feature of Figure \ref{fig:sflaw1} is that it predicts very strong variation in specific star formation rate with metallicity for CO and total gas, but not for H$_2$. In the low metallicity galaxies where we would like to test this prediction it is difficult to measure either the total gas mass or the H$_2$ mass, because we lack a reliable H$_2$ tracer except in a few limited cases (see Section \ref{futurework} for further discussion). We can, however, measure CO luminosities and thereby estimate masses of gas that are CO-dominated. As shown in Figure \ref{fig:sflaw1}, we predict that the specific star formation rate for CO should be much higher in the lower metallicity systems.

To test this prediction, we compile galaxy-integrated CO luminosities, $L_{\rm CO}$, and star formation rates, SFRs, from the literature. These measurements cover a wide range of metallicity from highly sub-Solar to super-Solar, and include both spirals and dwarfs. The data come from a variety of sources, summarized in
Table \ref{tab:data}. Whenever they are available, we give preference to $L_{\rm CO}$ derived
from integrating complete single-dish maps of a galaxy, but many of the
data still come from sparse sampling. For SFRs, we draw heavily from the
recent set of galaxy-integrated measurements by \citet{calzetti10a}; these have
the advantage of being computed in a uniform way and include an IR
contribution to correct for extinction. Both $L_{\rm CO}$ and the SFR are
luminosity-like quantities, so the ratio of the two is independent of
distance. We also draw metallicities for each target from the literature.
When available, we give preference to the recent compilation by \citet{moustakas10a}, using the average of their two characteristic metallicities. For targets not studied by \citeauthor{moustakas10a}, we
use values from the compilations by \citet{marble10a}, \citet{calzetti10a}, and \citet{engelbracht08a} and a variety of literature
sources.

\begin{deluxetable*}{ccccccc}
\tablecaption{Data compilation\label{tab:data}}
\tablehead{
\colhead{Galaxy} &
\colhead{$\log L_{\rm CO}$ [K km s${-1}$ pc$^2$]} &
\colhead{CO reference} &
\colhead{$\log \dot{M}_*$ [$\msun$ yr$^{-1}]$} &
\colhead{SFR reference} &
\colhead{$12 + \log({\rm O}/{\rm H})$} &
\colhead{Metallicity reference}
}
\startdata
SMC  &  $5.2 \pm 0.2$  &  M06  &  $-1.3 \pm 0.2$  &  W04  &  $8.0 \pm 0.2$  &  D84; MA10 \\
LMC  &  $6.5 \pm 0.1$  &  F08  &  $-0.7 \pm 0.2$  &  H09  &  $8.3 \pm 0.3$  &  D84; MA10 \\
IC10  &  $6.3 \pm 0.2$  &  L06  &  $-1.0 \pm 0.5$  &  L06  &  $8.2 \pm 0.2$  &  L79; L03 \\
M33  &  $7.6 \pm 0.1$  &  H04  &  $ 0.0 \pm 0.5$  &  H04  &  $8.3 \pm 0.2$  &  R08 \\
IIZw40  &  $6.2 \pm 0.3$  &  T98  &  $-0.2 \pm 0.3$  &  C10  &  $8.1 \pm 0.3$  &  E08; C10 \\
NGC1569  &  $5.1 \pm 0.3$  &  T98  &  $-0.6 \pm 0.3$  &  C10  &  $8.1 \pm 0.2$  &  M97 \\
NGC2537  &  $5.5 \pm 0.3$  &  T98  &  $-1.1 \pm 0.3$  &  C10  &  $8.4 \pm 0.3$  &  MA10 \\
NGC4449  &  $6.8 \pm 0.3$  &  Y95  &  $-0.5 \pm 0.3$  &  C10  &  $8.3 \pm 0.2$  &  M97 \\
NGC5253  &  $5.8 \pm 0.3$  &  T98  &  $-0.3 \pm 0.3$  &  C10  &  $8.2 \pm 0.3$  &  MA10 \\
NGC6822  &  $5.1 \pm 0.2$  &  I97  &  $-2.0 \pm 0.3$  &  C10  &  $8.4 \pm 0.3$  &  MO10 \\
NGC0628  &  $8.3 \pm 0.1$  &  L09  &  $-0.2 \pm 0.2$  &  C10  &  $8.8 \pm 0.3$  &  MO10 \\
NGC0925  &  $7.6 \pm 0.1$  &  L09  &  $-0.3 \pm 0.2$  &  C10  &  $8.6 \pm 0.3$  &  MO10 \\
NGC1482  &  $8.8 \pm 0.3$  &  Y95  &  $ 0.5 \pm 0.2$  &  C10  &  $8.6 \pm 0.4$  &  MO10 \\
NGC2146  &  $9.3 \pm 0.3$  &  Y95  &  $ 0.9 \pm 0.2$  &  C10  &  $8.7 \pm 0.3$  &  E08; C10 \\
NGC2403  &  $7.1 \pm 0.3$  &  Y95  &  $-0.4 \pm 0.2$  &  C10  &  $8.6 \pm 0.2$  &  MO10 \\
NGC2841  &  $8.4 \pm 0.1$  &  L09  &  $ 0.1 \pm 0.2$  &  C10  &  $9.0 \pm 0.3$  &  MO10 \\
NGC2782  &  $9.0 \pm 0.3$  &  Y95  &  $ 0.7 \pm 0.2$  &  C10  &  $8.6 \pm 0.4$  &  E08; C10 \\
NGC2798  &  $8.8 \pm 0.3$  &  Y95  &  $ 0.5 \pm 0.2$  &  C10  &  $8.7 \pm 0.3$  &  MO10 \\
NGC2976  &  $7.0 \pm 0.1$  &  L09  &  $-1.0 \pm 0.2$  &  C10  &  $8.7 \pm 0.3$  &  MO10 \\
NGC2903  &  $8.8 \pm 0.1$  &  H03  &  $ 0.3 \pm 0.2$  &  C10  &  $8.9 \pm 0.3$  &  MA10 \\
NGC3034  &  $8.8 \pm 0.3$  &  Y95  &  $ 0.9 \pm 0.2$  &  C10  &  $8.8 \pm 0.3$  &  MO10 \\
NGC3077  &  $6.0 \pm 0.3$  &  T98  &  $-1.0 \pm 0.2$  &  C10  &  $8.6 \pm 0.3$  &  MA10 \\
NGC3079  &  $9.4 \pm 0.3$  &  Y95  &  $ 0.5 \pm 0.2$  &  C10  &  $8.6 \pm 0.4$  &  E08; C10 \\
NGC3184  &  $8.4 \pm 0.1$  &  L09  &  $-0.5 \pm 0.2$  &  C10  &  $8.9 \pm 0.3$  &  MO10 \\
NGC3198  &  $7.9 \pm 0.1$  &  L09  &  $-0.0 \pm 0.2$  &  C10  &  $8.8 \pm 0.3$  &  MO10 \\
NGC3310  &  $8.2 \pm 0.3$  &  Y95  &  $ 0.9 \pm 0.2$  &  C10  &  $8.2 \pm 0.4$  &  E08; C10 \\
NGC3351  &  $8.2 \pm 0.1$  &  L09  &  $-0.2 \pm 0.2$  &  C10  &  $8.9 \pm 0.3$  &  MO10 \\
NGC3368  &  $8.3 \pm 0.3$  &  Y95  &  $-0.4 \pm 0.2$  &  C10  &  $9.0 \pm 0.3$  &  MA10 \\
NGC3521  &  $8.8 \pm 0.1$  &  L09  &  $ 0.1 \pm 0.2$  &  C10  &  $8.8 \pm 0.3$  &  MO10 \\
NGC3628  &  $9.2 \pm 0.3$  &  Y95  &  $ 0.3 \pm 0.2$  &  C10  &  $9.0 \pm 0.3$  &  MA10 \\
NGC3627  &  $9.0 \pm 0.1$  &  H03  &  $ 0.2 \pm 0.2$  &  C10  &  $8.7 \pm 0.3$  &  MO10 \\
NGC3938  &  $8.5 \pm 0.1$  &  H03  &  $-0.1 \pm 0.2$  &  C10  &  $8.7 \pm 0.3$  &  E08; C10 \\
NGC4194  &  $8.9 \pm 0.3$  &  Y95  &  $ 1.1 \pm 0.2$  &  C10  &  $8.7 \pm 0.4$  &  E08; C10 \\
NGC4214  &  $6.1 \pm 0.1$  &  L09  &  $-1.0 \pm 0.2$  &  C10  &  $8.2 \pm 0.2$  &  T98 \\
NGC4254  &  $9.9 \pm 0.3$  &  Y95  &  $ 1.3 \pm 0.2$  &  C10  &  $8.8 \pm 0.3$  &  MO10 \\
NGC4321  &  $9.2 \pm 0.1$  &  H03  &  $ 0.4 \pm 0.2$  &  C10  &  $8.8 \pm 0.3$  &  MO10 \\
NGC4450  &  $8.9 \pm 0.3$  &  Y95  &  $-0.2 \pm 0.2$  &  C10  &  $8.9 \pm 0.4$  &  C10; MA10 \\
NGC4536  &  $8.6 \pm 0.3$  &  Y95  &  $ 0.3 \pm 0.2$  &  C10  &  $8.6 \pm 0.4$  &  MO10 \\
NGC4569  &  $8.8 \pm 0.1$  &  H03  &  $-0.1 \pm 0.2$  &  C10  &  $8.9 \pm 0.4$  &  E08; C10 \\
NGC4579  &  $9.0 \pm 0.3$  &  Y95  &  $ 0.2 \pm 0.2$  &  C10  &  $9.0 \pm 0.4$  &  C10; MA10 \\
NGC4631  &  $8.5 \pm 0.3$  &  Y95  &  $ 0.4 \pm 0.2$  &  C10  &  $8.4 \pm 0.3$  &  MO10 \\
NGC4725  &  $9.1 \pm 0.3$  &  Y95  &  $ 0.0 \pm 0.2$  &  C10  &  $8.7 \pm 0.4$  &  MO10 \\
NGC4736  &  $7.9 \pm 0.1$  &  L09  &  $-0.4 \pm 0.2$  &  C10  &  $8.7 \pm 0.4$  &  MO10 \\
NGC4826  &  $8.1 \pm 0.1$  &  H03  &  $-0.5 \pm 0.2$  &  C10  &  $8.9 \pm 0.3$  &  MO10 \\
NGC5033  &  $9.3 \pm 0.1$  &  H03  &  $ 0.1 \pm 0.3$  &  K03  &  $8.7 \pm 0.3$  &  MO10 \\
NGC5055  &  $8.9 \pm 0.1$  &  L09  &  $ 0.1 \pm 0.2$  &  C10  &  $8.9 \pm 0.4$  &  MO10 \\
NGC5194  &  $9.2 \pm 0.1$  &  H03  &  $ 0.4 \pm 0.2$  &  C10  &  $9.0 \pm 0.3$  &  MO10 \\
NGC5236  &  $8.9 \pm 0.1$  &  Y95  &  $ 0.4 \pm 0.2$  &  C10  &  $9.0 \pm 0.3$  &  MO10 \\
NGC5713  &  $9.1 \pm 0.3$  &  Y95  &  $ 0.6 \pm 0.2$  &  C10  &  $8.7 \pm 0.4$  &  MO10 \\
NGC5866  &  $8.1 \pm 0.3$  &  Y95  &  $-0.6 \pm 0.2$  &  C10  &  $8.7 \pm 0.4$  &  C10; MA10 \\
NGC5953  &  $9.0 \pm 0.3$  &  Y95  &  $ 0.4 \pm 0.2$  &  C10  &  $8.7 \pm 0.4$  &  E08; C10 \\
NGC6946  &  $8.8 \pm 0.1$  &  L09  &  $ 0.5 \pm 0.2$  &  C10  &  $8.8 \pm 0.3$  &  MO10 \\
NGC7331  &  $9.0 \pm 0.1$  &  L09  &  $ 0.4 \pm 0.2$  &  C10  &  $8.8 \pm 0.3$  &  MO10 \\
\enddata
\tablecomments{
C10 = \citet{calzetti10a}; 
D84 = \citet{dufour84a}; 
 E08 = \citet{engelbracht08a}; 
 F08 = \citet{fukui08a}; 
H03 = \citet{helfer03a}; 
H04 = \citet{heyer04b}; 
H09 = \citet{harris09a}; 
I97 = \citet{israel97a}; 
K03 = \citet{kennicutt03a}; 
L79 = \citet{lequeux79a};
 L03 = \citet{lee03a}; 
L06 = \citet{leroy06a}; 
L09 = \citet{leroy09a}; 
M97 = \citet{martin97a}; 
M06 = \citet{mizuno06a};
MA10 = \citet{marble10a}; 
MO10 = \citet{moustakas10a}
R08 = \citet{rosolowsky08b}; 
 T98 = \citet{taylor98a}; 
 W04 = \citet{wilke04a}; 
 Y95 = \citet{young95a}
}
\end{deluxetable*}

\begin{figure}
\epsscale{1.2}
\plotone{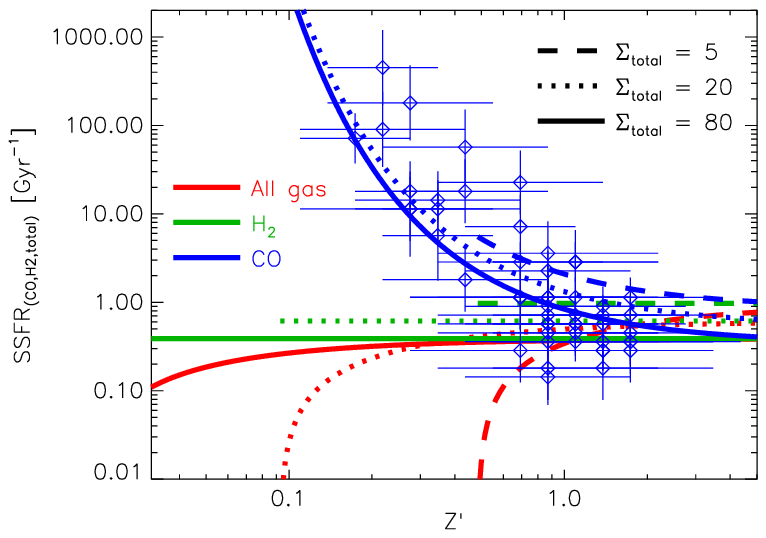}
\epsscale{1.0}
\caption{
\label{fig:sflaw2}
Predicted gaseous specific star formation rate ${\rm SSFR}_{(\rm CO,H_2,total)} = \dot{\Sigma}_*/\Sigma_{(\rm CO,H_2,total)} = \dot{M}_*/M_{(\rm CO,H_2,total)}$ as a function metallicity. As in Figure \ref{fig:sflaw1}, the red lines indicate the specific star formation rate considering the total gas mass, green lines indicate the specific star formation rate for H$_2$, while blue lines indicate the specific star formation rate for gas where the carbon is CO rather than C~\textsc{ii}. Dashed lines correspond to a total gas surface density $\Sigma_{\rm total}=5$ $\msun$ pc$^{-2}$, dotted lines to $\Sigma_{\rm total}=20$ $\msun$ pc$^{-2}$, and solid lines to $\Sigma_{\rm total}=80$ $\msun$ pc$^{-2}$. Note that these are all total gas surface densities, rather than surface densities of H$_2$ or CO as on the $x$ axis in Figure \ref{fig:sflaw1}. Also note that we only show values for ${\rm SSFR}_{(\rm CO)}$ and ${\rm SSFR}_{(\rm H_2)}$ for metallicities and surface densities that are high enough for$f_{\rm H_2}$ and $f_{\rm CO}$ to be non-zero.
}
\end{figure}

To convert observed CO $(1-0)$ luminosities to masses of gas traced by CO,
we adopt a conversion factor $2 \times 10^{20}$ H$_2$ molecules cm$^{-2} / (\mbox{K km s}^{-1})^{-1}$
\citep{abdo10b, blitz07a, draine07a, heyer09a}, so the
mass is
\begin{equation}
M_{\rm CO} = 4\times 10^{20} \mu_{\rm H} \left(\frac{L_{\rm CO}}{\mbox{K km s}^{-1}\mbox{ cm}^2}\right),
\end{equation}
where $\mu_{\rm H}$ is the mean mass per H nucleus. This is equivalent to
\begin{equation}
\frac{M_{\rm CO}}{\msun} = \alpha_{\rm CO} \left(\frac{L_{\rm CO}}{\mbox{K km s}^{-1}\mbox{ pc}^2}\right)
\end{equation}
with $\alpha_{\rm CO}=4.4$. Thus, we
effectively assume a Milky Way conversion factor for CO-emitting gas. It is important to note that our $X$ factor represents the conversion from CO luminosity to mass of gas where the carbon is predominantly CO, and {\it not} the conversion from CO luminosity to total mass of gas where the hydrogen is H$_2$. These concepts are often not clearly distinguished in the literature. The
approximately constant conversion factor derived from virial mass
measurements of extragalactic clouds \citep{blitz07a, bolatto08a} (which include some but not all of the H$_2$ that is not associated with CO)
motivates this assumption, though there may still be changes in $\alpha_{\rm CO}$ of
CO-emitting at the factor of 2 level across the range of metallicities that
we study. The sense of these would be to increase $\alpha_{\rm CO}$, and to decrease the
SFR-to-$M_{\rm CO}$ ratio, moving points down in Figure \ref{fig:sflaw2}. Regardless of
systematic effects, the y-axis in Figure \ref{fig:sflaw2} should be very close to the ratio of
ionizing photon rate to CO luminosity.

To convert measured oxygen abundances to metallicities relative to Milky
Way we adopt \citep{caffau08a}
\begin{equation}
\log Z' = [12 + \log({\rm O}/{\rm H})] - 8.76.
\end{equation}
As emphasized by \citet{kewley08a} and \citet{moustakas10a}, the adopted calibration has a large influence on the
metallicities derived from measurements of strong optical lines. Therefore
the overall normalization of the metallicities in Figure \ref{fig:sflaw2} must be considered uncertain by at least $\sim 0.1
- 0.2$ dex, though the internal ordering is likely to be relatively robust.

A final complication is that most of the literature data we have gathered consists of galaxy-integrated values, rather than local values. We therefore do not have surface densities, and we must adopt characteristic values in order to compute the relationship between total gas, molecular gas, and CO luminosity. Fortunately, the results for the specific star formation rate for CO are not particularly sensitive to this assumption, as shown below.

We plot the literature data against our predictions in Figure \ref{fig:sflaw2}. As the plot shows, there is a clear a correlation between SSFR$_{\rm CO}$ and metallicity that agrees within the uncertainties with what one expects if star formation follows H$_2$ rather than total gas or CO. Though the data are sparse with significant
scatter, the difference appears to be more than 2 orders of magnitude in
the SFR-to-CO ratio over about an order of magnitude in metallicity. The
shape of this trend agrees well with our theoretical predictions. Given the uncertainties in estimating several of the
parameters shown here, we believe this constitutes a reasonable first-check
on the model. Future improvements to the measured
dust-to-gas ratios, improved estimates of $\alpha_{\rm CO}$ from CO emitting gas, and
observations of CO from larger samples of low-metallicity galaxies will
improve the accuracy of the observed trend and allow more stringent tests.

The elevated value of SSFR$_{\rm CO}$ in dwarf galaxies and the outer parts of spirals relative to the inner parts of spirals has been pointed out before \citep{young96a, leroy06a, leroy07b, gardan07a}, but it was unclear if this was due to change in the star formation process or a change in the CO to H$_2$ ratio combined with star formation following H$_2$ rather than CO. \citet{pelupessy09a} argued that the elevated value of SSFR$_{\rm CO}$ could be explained if the latter were true, but they simply assumed that star formation was correlated with H$_2$, rather than explaining the correlation from first principles. Here we have provided the missing explanation, and Figure \ref{fig:sflaw2} demonstrates that a model based on it can quantitatively reproduce the observations.

\subsection{Predictions for Future Observations}
\label{futurework}

Figures \ref{fig:sflaw1} and \ref{fig:sflaw2} also contain clear predictions for observations. At present it is extremely difficult to measure either total gas masses or H$_2$ masses in galaxies with low metallicity, because CO breaks down as a tracer of H$_2$ in these environments, and because the H$_2$ itself does not emit. Observations are available for only a few galaxies based on using proxies other than CO for the H$_2$ \citep[e.g.][]{leroy07a}. However, future dust observations with facilities such as Herschel and ALMA will make it easier to obtain H$_2$ masses, and thus total gas masses, for more galaxies. Our work contains a clear prediction: for these galaxies, the specific star formation rate with respect to H$_2$ mass should be essentially independent of metallicity, while the specific star formation rate for the total gas mass will behave in the opposite sense as for CO: low metallicity galaxies will have lower $\mbox{SSFR}_{\rm total}$, even as they have higher $\mbox{SSFR}_{\rm CO}$.

With resolved observations an additional test becomes possible. Figure \ref{fig:sflaw1} shows that $\mbox{SSFR}_{\rm total}$ is essentially flat at high surface densities, but turns down sharply at surface densities below a metallicity-dependent value. Such a drop is seen below $\Sigma_{\rm total} \la 10$ $\msun$ pc$^{-2}$ in observations of Solar metallicity galaxies \citet{bigiel08a}. Models in which the star formation rate depends on global gravitational instability or similar phenomena that do not care about gas cooling or chemistry predict that the surface density at which the total gas specific star formation rate drops should not vary with metallicity \citep[e.g.][]{li06c}. In contrast, our work here suggests that it should scale roughly inversely with metallicity; \citet{schaye04a}, using a thermally-based model that anticipates some of this work, makes a similar prediction (his equation 25), although the metallicity-dependence in his model is weaker than in ours. Resolved measurements of the total gas surface density in nearby galaxies should be able to test which prediction is correct: does $\mbox{SSFR}_{\rm total}$ always change sharply at $\sim 10$ $\msun$ pc$^{-2}$, or is that value metallicity-dependent? Preliminary results indicate appear consistent with a dependence on metallicity \citep{fumagalli10a}, but a more systematic survey is needed.

\section{Summary}
\label{summary}

In this paper we consider which phase of the interstellar medium should correlate best with the star formation rate, and why. Our work extends the previous examination of this problem by \citet{schaye04a}. We use chemical and thermal models of interstellar clouds to show that the star formation rate is expected to correlate most closely with the molecular hydrogen content of a galaxy. In order to undergo runaway gravitational collapse to form stars, gas must be able to reach low temperatures and therefore low Bonnor-Ebert masses. Its ability to do so, however, is impaired by the interstellar radiation field, which heats the gas. Only in regions where the ISRF is sufficiently excluded by extinction can star formation occur. We find that such regions are also the regions where the gas is expected to be predominantly H$_2$ rather than H~\textsc{i}, and for this reason the star formation rate correlates with, but is not caused by, the H~\textsc{i} to H$_2$ transition. In contrast, the chemical makeup has relatively little effect on the ability of the gas to cool. All of these results are robust against a very wide range of variation in metallicity, radiation field, or other properties of the galactic environment.

That star formation correlates with H$_2$ rather than either total gas mass or CO mass has strong observational implications \citep[see also][]{pelupessy09a}. The fraction of H$_2$ gas where the carbon is in the form of C~\textsc{ii} rather than CO is a strong function of metallicity. Galaxies with low metallicity tend to have large masses of H$_2$ where there is fairly little CO. If this material is able to form stars, as we predict, then the star formation rate per unit CO mass should be very large in low metallicity galaxies. We see that precisely this phenomenon is found in observed galaxies. Finally, we note that the fraction of the total gas mass where the hydrogen is H$_2$ rather than H~\textsc{i} is also a strongly increasing function of metallicity. Since star formation correlates with H$_2$, we predict that the star formation rate per unit total gas mass should be small in low metallicity galaxies. This prediction can be used to test our calculations.

\acknowledgements We thank E.~Ostriker, M.~Reiter, and an anonymous referee for comments that improved the manuscript. This work was supported by an Alfred P. Sloan Fellowship (MRK); NSF grants AST-0807739 (MRK), CAREER-0955300 (MRK), and AST-0908553 (CFM); and NASA through Astrophysics Theory and Fundamental Physics grant NNX09AK31G (MRK and CFM), through a Spitzer Space Telescope Theoretical Research Program grant (MRK and CFM), and through Hubble Fellowship grant HST-HF-51258.01-A awarded by the Space Telescope Science Research Institute, which is operated by the Association of Universities for Research in Astronomy, Inc., for NASA, under contract NAS 5-26555 (AKL). MRK and AKL thank the Aspen Center for Physics, where much of the work for this paper was performed.

\begin{appendix}

\setcounter{figure}{0}

\section{Parameter Choices}
\label{app:paramdis}

All molecular and atomic information, including level energies, Einstein coefficients, and collision rate coefficients, is taken from the Leiden Atomic and Molecular Database \citep{schoier05a}. Other input parameters are gathered from a variety of sources, and our fiducial values and references are given in Table \ref{tab:param}. Most of these are straightforward, and where a quantity has been observed in the Milky Way, we extrapolate to other galaxies by assuming that element abundances and dust to gas ratios are simply proportional to metallicity. Parameters that are not directly observed or extrapolated we discuss in the remainder of this Appendix.

\subsection{Dust-Gas Thermal Exchange Rate Coefficients} 

For the dust-gas thermal exchange rate coefficient $\alpha_{\rm dg}$, we adopt a standard value of $3.8\times 10^{-34}$ erg cm$^3$ K$^{-3/2}$ in H$_2$-dominated gas in the Milky Way \citep{goldsmith01a}, and we assume that this will also hold in galaxies of similar metallicity. We must extrapolate this both to lower metallicity galaxies, and to predominantly H~\textsc{i} gas. For the former, we assume that the total surface area of grains is proportional to the metal abundance. For the latter, we must account for both the change in the number of particles and the masses of the individual particles, which alters their speed. We assume that the accommodation coefficient is equal for H, H$_2$, and He. (More accurate approximations are possible if the grain size distribution is known, e.g.~\citet{hollenbach79a}, but given the uncertainties in how size distributions vary from galaxy to galaxy and the relative lack of importance of this effect, we omit this complication.) With this assumption, the dust-gas energy exchange rate provided by gaseous members of species $X$ is proportional to $n_X/m_X^{1/2}$, where $n_X$ is the number density of members of that species and $m_X$ is the mass of that species. Thus in regions of atomic and molecular hydrogen respectively we have
\begin{eqnarray}
\alpha_{\rm dg, HI} & \propto & \frac{n_{\rm HI}}{m_{\rm H}^{1/2}} + \frac{n_{\rm He}}{m_{\rm He}^{1/2}} \\
\alpha_{\rm dg, H_2} & \propto & \frac{n_{\rm H_2}}{m_{\rm H_2}^{1/2}} + \frac{n_{\rm He}}{m_{\rm He}^{1/2}},
\end{eqnarray}
where the constants of proportionality are the same in each case. In a predominantly atomic region $n_{\rm HI} = n_{\rm H}$, while in a predominantly molecular region $n_{\rm H_2} = n_{\rm H}/2$; in both cases $n_{\rm He} \approx n_{\rm H}/10$. Similarly, $m_{\rm He} = 4 m_{\rm H}$ and $m_{\rm H_2} = 2 m_{\rm H}$. Combining the dependence on metallicity with that on chemical phase, we arrive at our final expressions for $\alpha_{\rm dg}$:
\begin{eqnarray}
\alpha_{\rm dg, H_2} & = & 3.8\times 10^{-33} Z'\mbox{ erg cm}^3\mbox{ K}^{-3/2} \\
\alpha_{\rm dg, HI} & = & 1.0\times 10^{-33} Z' \mbox{ erg cm}^3\mbox{ K}^{-3/2}.
\end{eqnarray}
For the radiation temperature, $T_{\rm rad}$, as discussed above we must select a single value to characterize the re-radiated infrared field within a cloud. We adopt $T_{\rm rad} = 8$ K for this, near the minimum temperature seen in ammonia observations of Galactic cold clouds \citep{jijina99a}. Both this choice and our choice of $\alpha_{\rm dg}$ have very little impact on our results due to the weak dust-gas coupling in the density range with which we are concerned.

\subsection{Opacities} 

Our calculations depend on the absorption cross section per H nucleus $\sigma_d$ to UV photons, and the visual extinction per unit hydrogen column $A_V/N_{\rm H}$. Note that $\sigma_d$ reflects only absorption, while $A_V$ includes scattering as well. These quantities depend on the extinction curve and vary between dense and diffuse environments even within a single galaxy at a single metallicity, and thus their exact values are uncertain by a factor of two. As a guide we examine the values given by the models of \citet{weingartner01a} for the Milky Way, Large Magellanic Cloud, and Small Magellanic Cloud.\footnote{\citet{draine03a} gives updated models for the Milky Way, but since these are not available for the LMC and SMC, we use the older models instead. The difference is at most tens of percent.} Once we scale the LMC and SMC models to the same total dust to gas ratio as the Milky Way models, we obtain values of $\sigma_d/10^{-21}\mbox{ cm}^{-2} = 0.7, 1.7, 2.0, 1.8$ and $(A_V/N_{\rm H})/ 10^{-22}\mbox{ mag cm}^{-2} = 4.5, 5.1, 3.6, 2.8$ for \citeauthor{weingartner01a}'s models for Milky Way sightlines with $R_V = 5.5$, Milky Way sightlines with $R_V = 3.1$, the average LMC sightline, and a sightline through the SMC bar, respectively. We therefore adopt intermediate values $\sigma_d=1.0\times 10^{-21} Z'$ cm$^2$ and $A_V/N_{\rm H} = 4.0\times 10^{-22} Z'$ mag cm$^2$. With these fiducial choices, the UV absorption is larger than the the visual extinction by a factor of $2.5/1.08=2.31$ (where the factor of 1.08 accounts for the conversion from magnitudes to true dimensionless units).

\subsection{Far Ultraviolet Radiation Field} 

The FUV radiation field, parameterized by $G_0'$, affects both gas temperature and chemistry. There is no unique value of $G_0'$ that characterizes all gas clouds across all galaxy types, and even within a single star-forming cloud the ambient UV field increases with time as stars form. As a fiducial choice we adopt $G_0'=1$, the Solar neighborhood value. In regions where star formation is ongoing the mean is likely larger, but we are interested in the initiation of star formation, so it seems prudent to select a value typical of regions before the onset of star formation. 

\begin{figure}
\plottwo{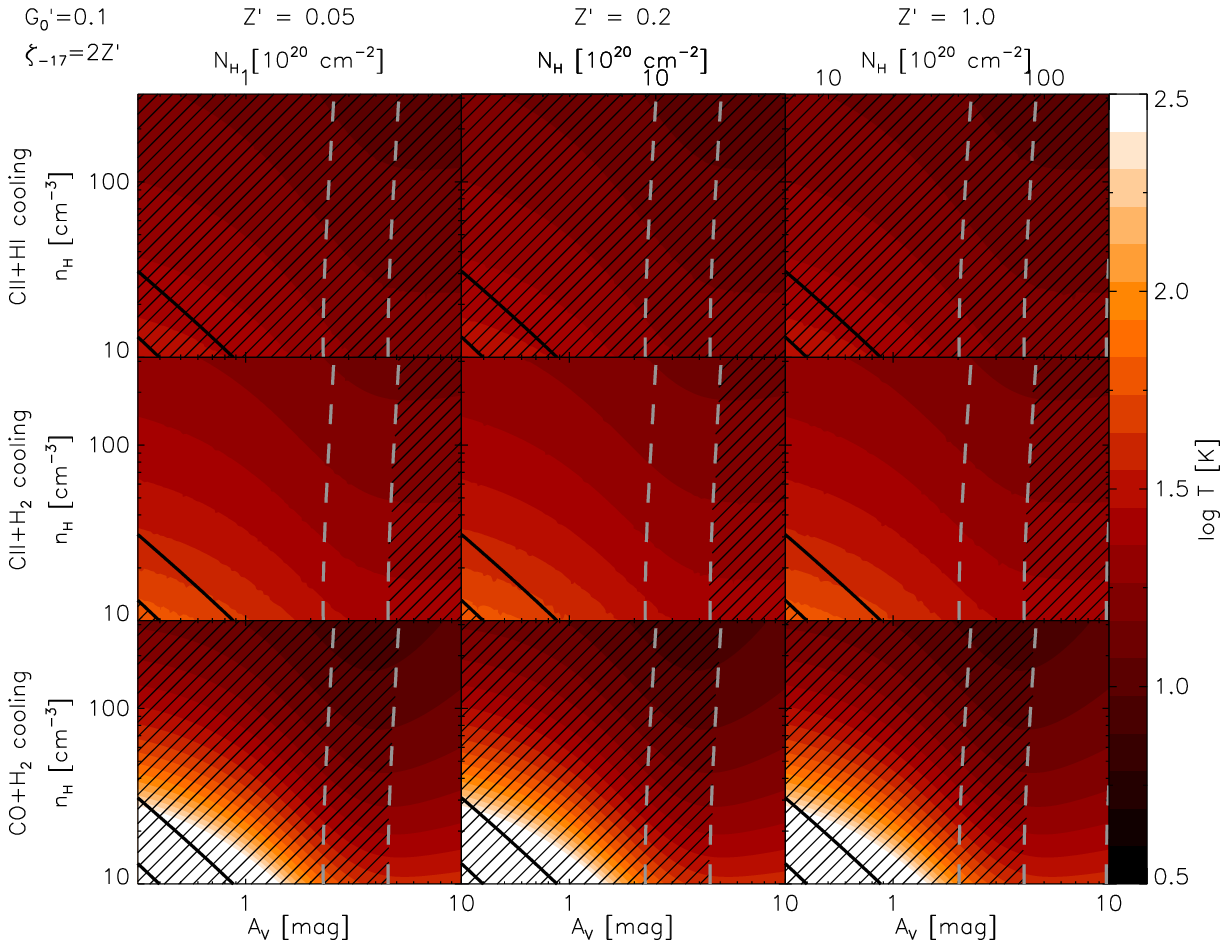}{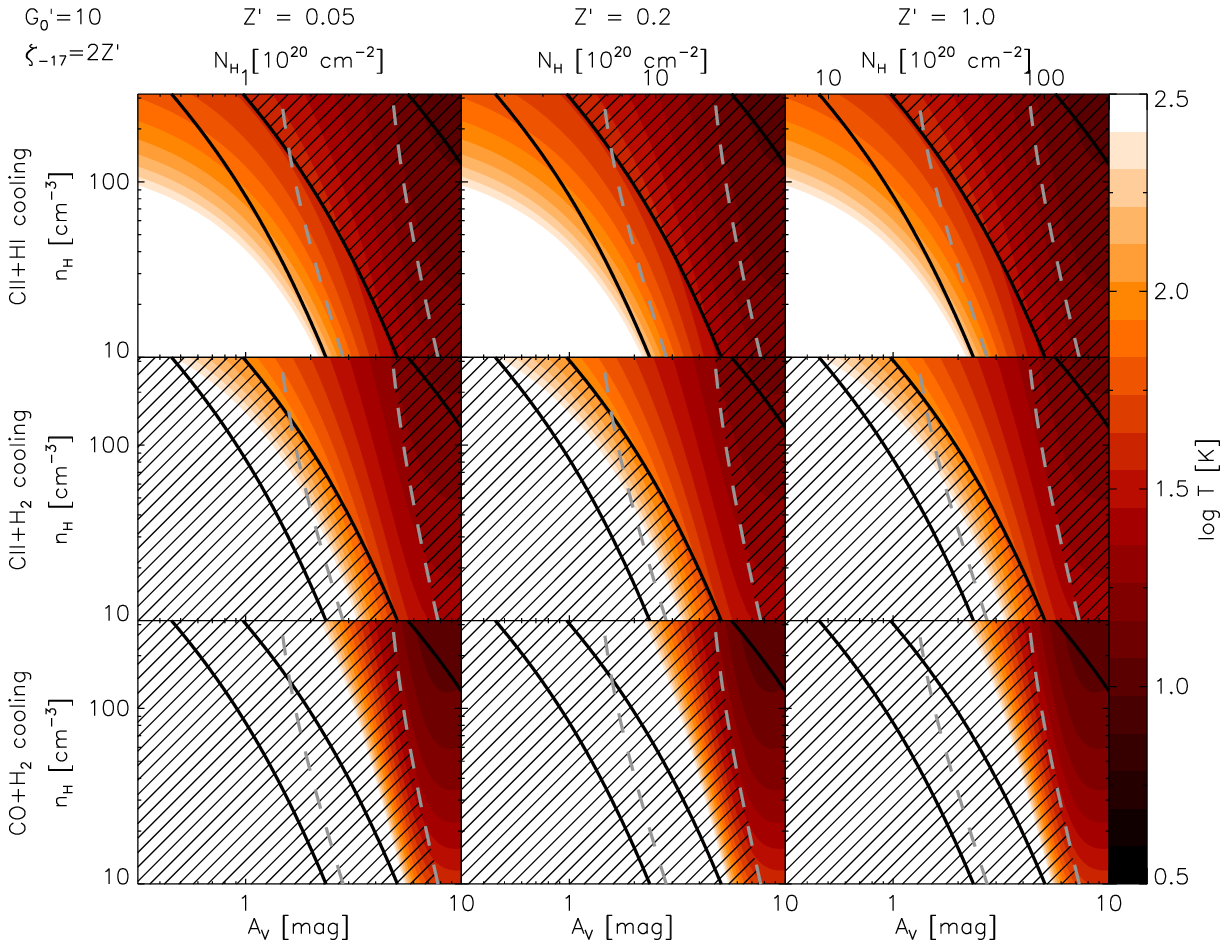}
\plottwo{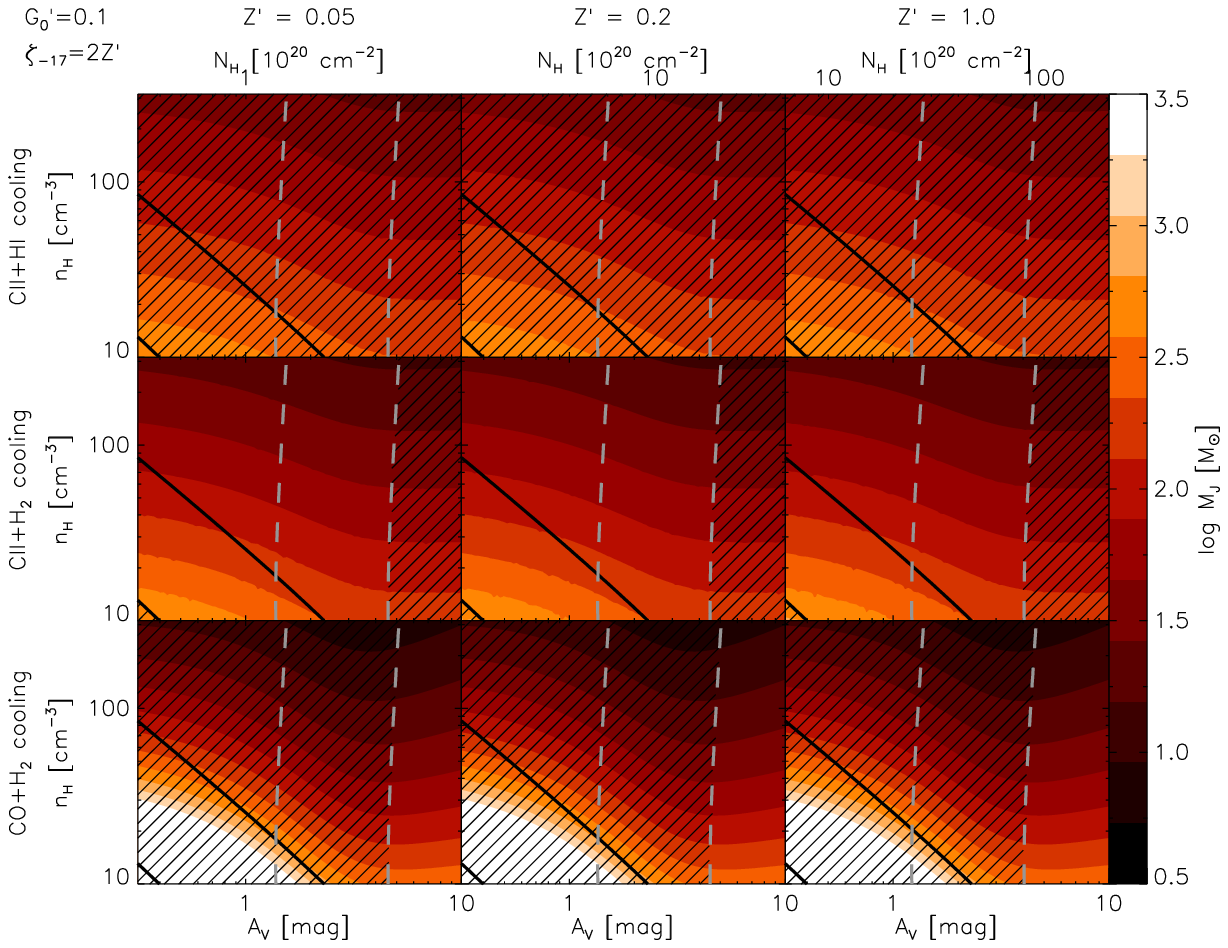}{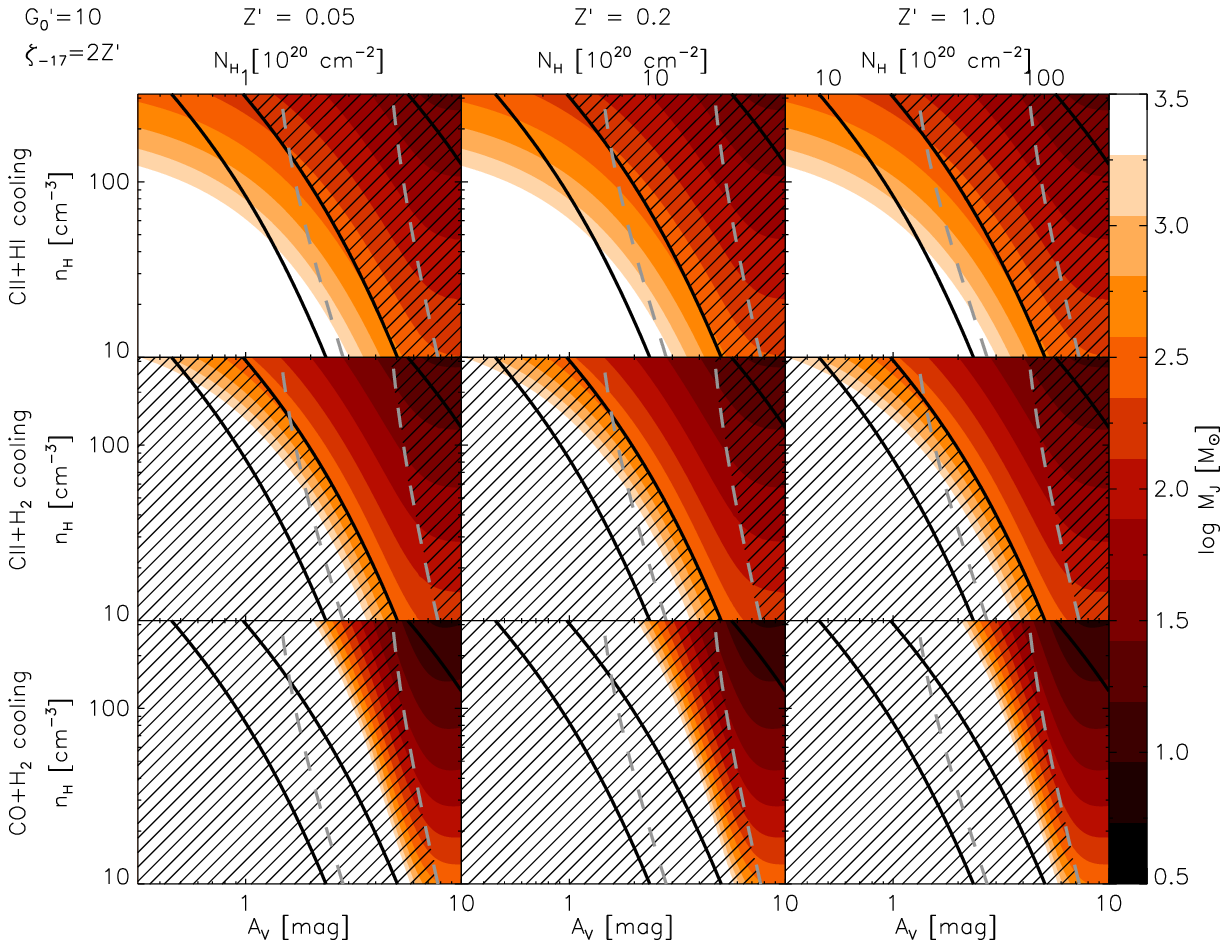}
\plottwo{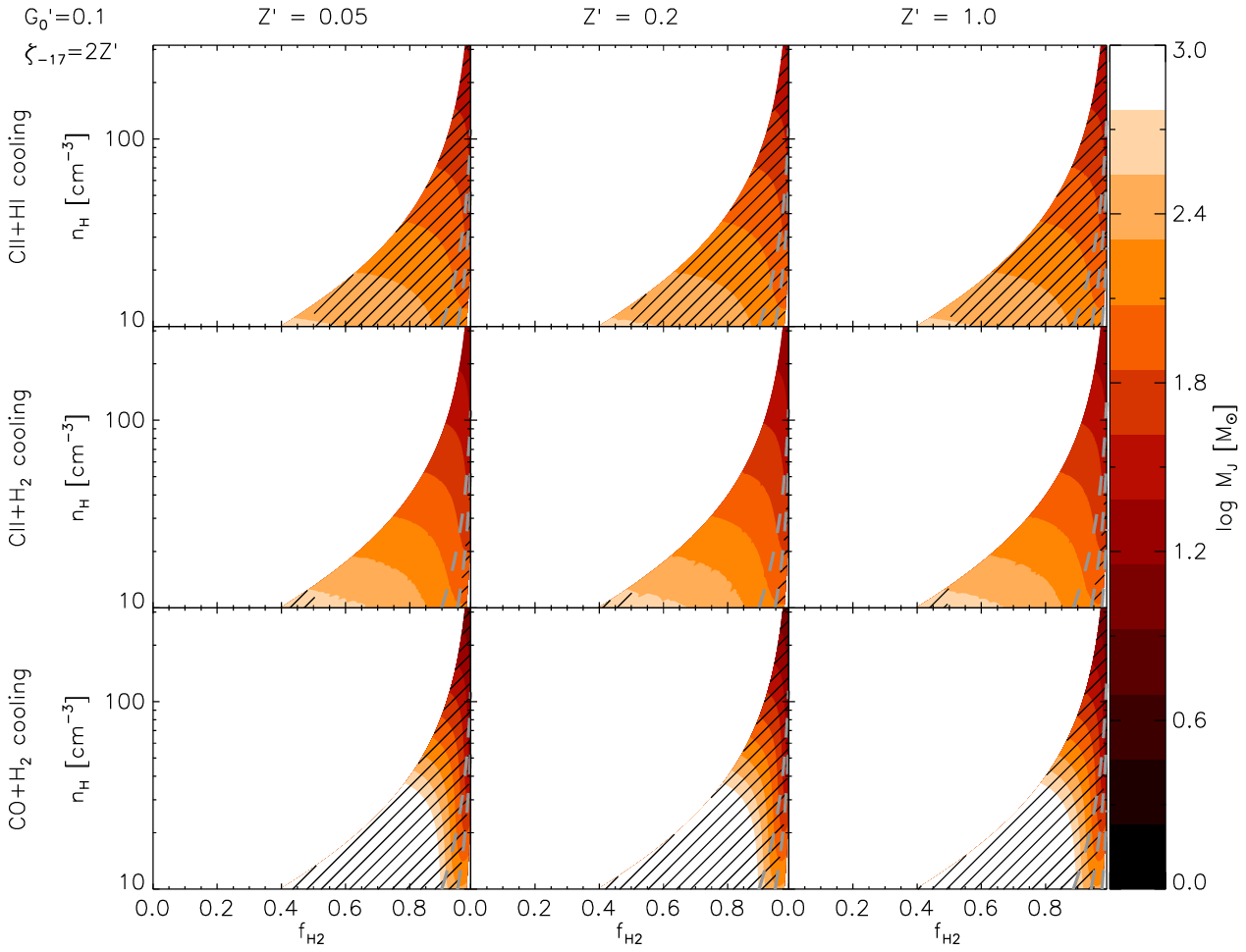}{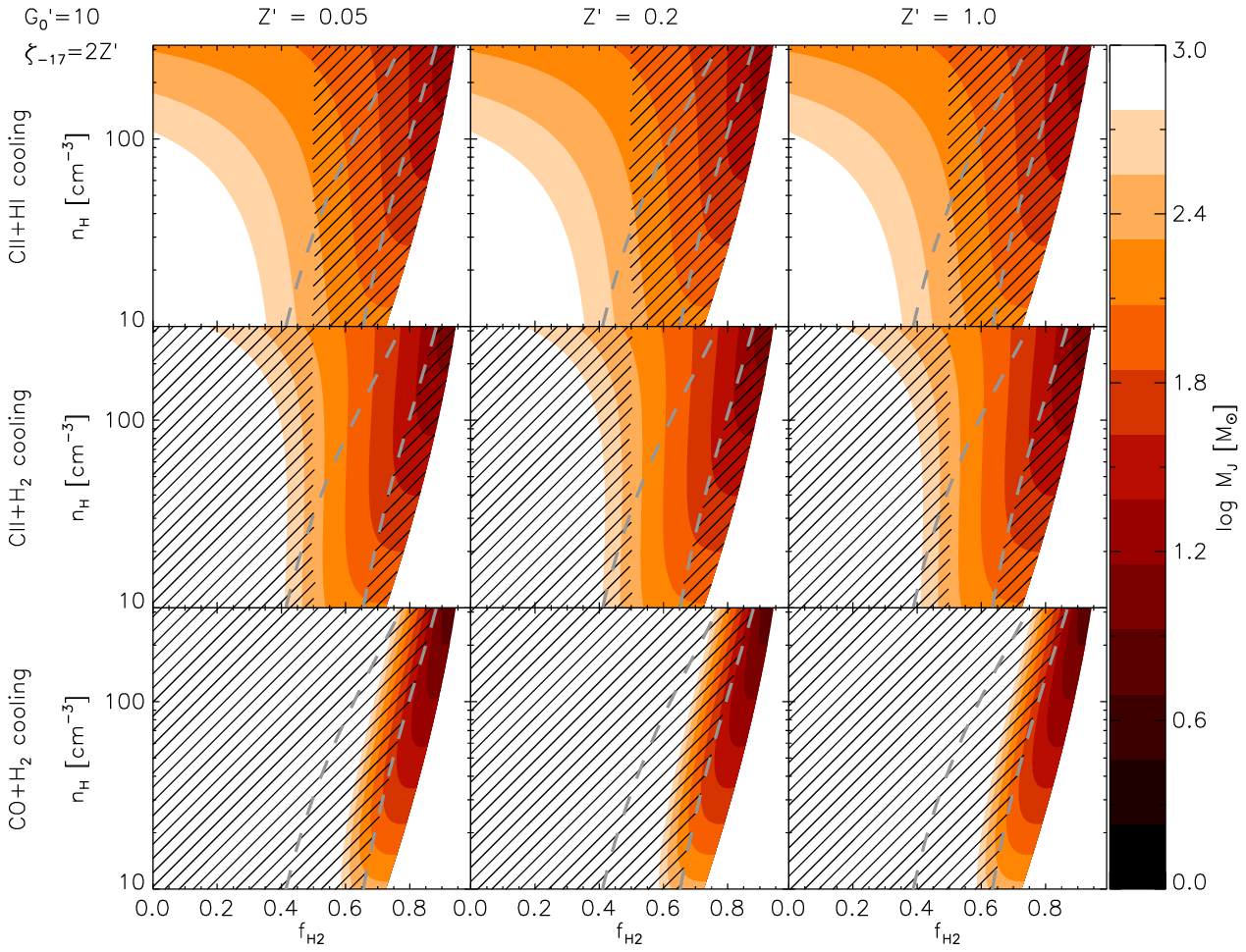}
\caption{
\label{fig:ggrid}
Left column: same as Figures \ref{fig:tempgrid}, \ref{fig:mjgrid}, and \ref{fig:mjh2}, but with $G_0' = 0.1$ instead of $G_0' = 1$. Right column: same as Figures \ref{fig:tempgrid}, \ref{fig:mjgrid}, and \ref{fig:mjh2}, but with $G_0' = 10$ instead of $G_0' = 1$.
}
\end{figure}

To test our sensitivity to this choice, we have recomputed our model grids with values of $G_0' = 0.1$ and $G_0' = 10$, and re-plotted Figures \ref{fig:tempgrid} -- \ref{fig:mjh2} of the main text in Figure \ref{fig:ggrid}. Not surprisingly, the H~\textsc{i} to H$_2$, C~\textsc{ii} to CO transitions, and warm to cold gas transitions all move to higher density and $A_V$ for higher $G_0'$, and to lower density and $A_V$ for lower $G_0'$. Nonetheless, over a factor of 100 range in $G_0'$, we will retain the excellent correlation between the H~\textsc{i} to H$_2$ transition and the drop in temperature from hundreds of K to $\sim 10$ K, along with the concomitant drop in the Bonnor-Ebert mass from several $1000$ $\msun$ or more to a few $\msun$. At all three values of $G_0'$, contours of constant $f_{\rm H_2}$ align closely with contours of constant Bonnor-Ebert mass. In the lower two panels, note that, contours of constant $M_{\rm BE}$ remain largely vertical, indicating that the Bonnor-Ebert mass drops systematically as the H$_2$ fraction increases. Thus our conclusions are robust against large variations in $G_0'$.

\subsection{Cosmic Rays}

The cosmic ray heating rate is uncertain in two ways. First, the energy yield per primary cosmic ray ionization $q_{\rm CR}$ is 6.5 eV in predominantly neutral H~\textsc{i} \citep{dalgarno72a, wolfire95a}. In H$_2$ the yield is uncertain. Both dissociative recombination of H$_2$ and excitation of its rotational and vibrational levels followed by collisional de-excitation provide extra channels for energy transfer from primary cosmic ray electrons to thermal motion, but the importance of these processes is likely density-dependent. Values given in the literature range from nearly the same energy yield as in H~\textsc{i} up to 20 eV per primary ionization \citep{glassgold73a, dalgarno99a}. We follow \citet{wolfire10a} in adopting an intermediate value $q_{\rm CR} = 12.25$ eV, but this should be regarded as uncertain by a factor of 2.

A similar uncertainty affects the cosmic ray ionization rate. For Milky Way-like galaxies we adopt $\zeta=2\times 10^{-17}$ s$^{-1}$ \citep{wolfire10a}, but the primary cosmic ray ionization rate even in the Milky Way is substantially uncertain, and may be higher than this value \citep{neufeld10a}. The cosmic ray intensity also varies with the star formation rate in galaxies \citep{abdo10a}. Thus in low metallicity galaxies, which also tend to have low star formation rates, the cosmic ray ionization rate is likely lower than the Milky Way value. The scaling is extremely uncertain; for lack of a better alternative, we simply take the cosmic ray ionization rate to scale with the metallicity.

\begin{figure}
\plottwo{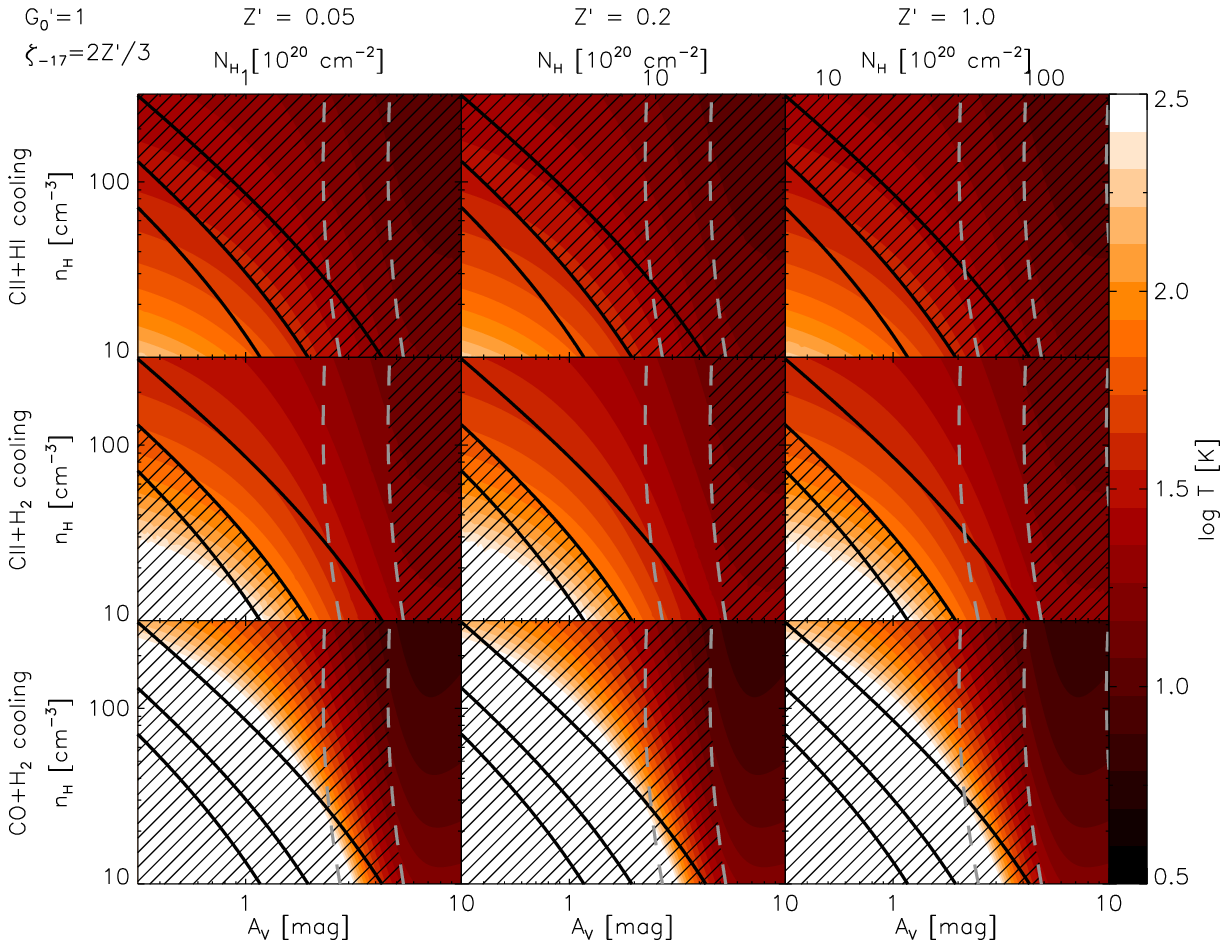}{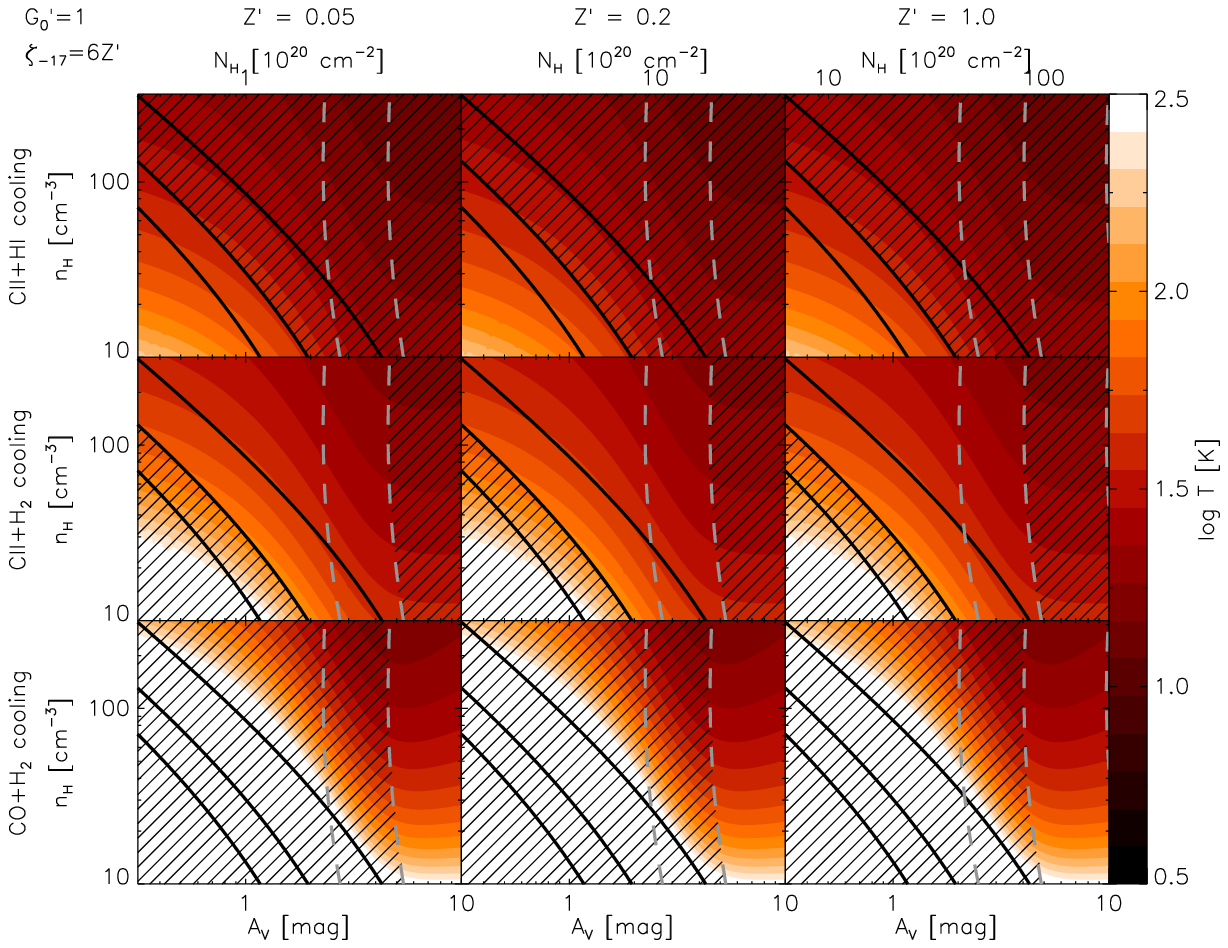}
\plottwo{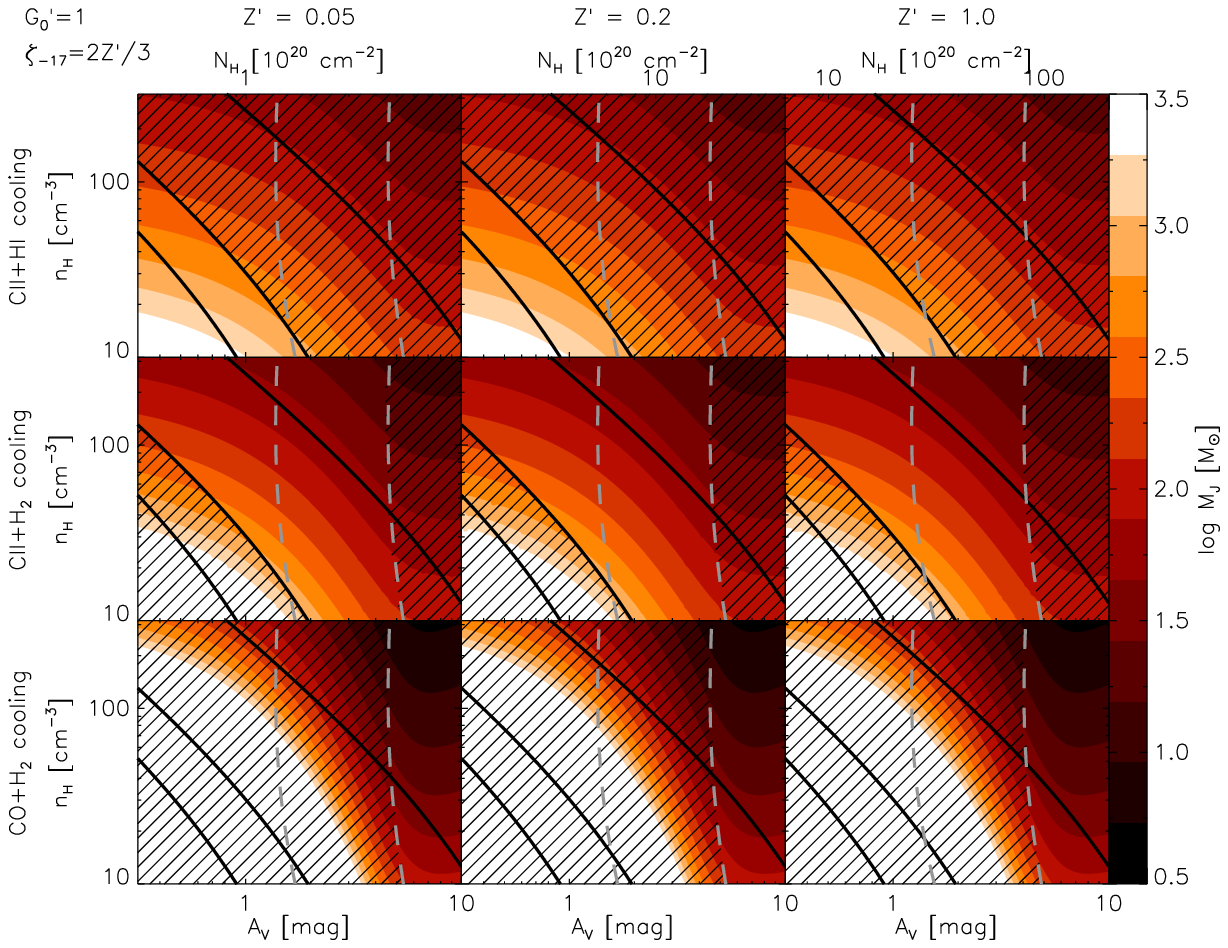}{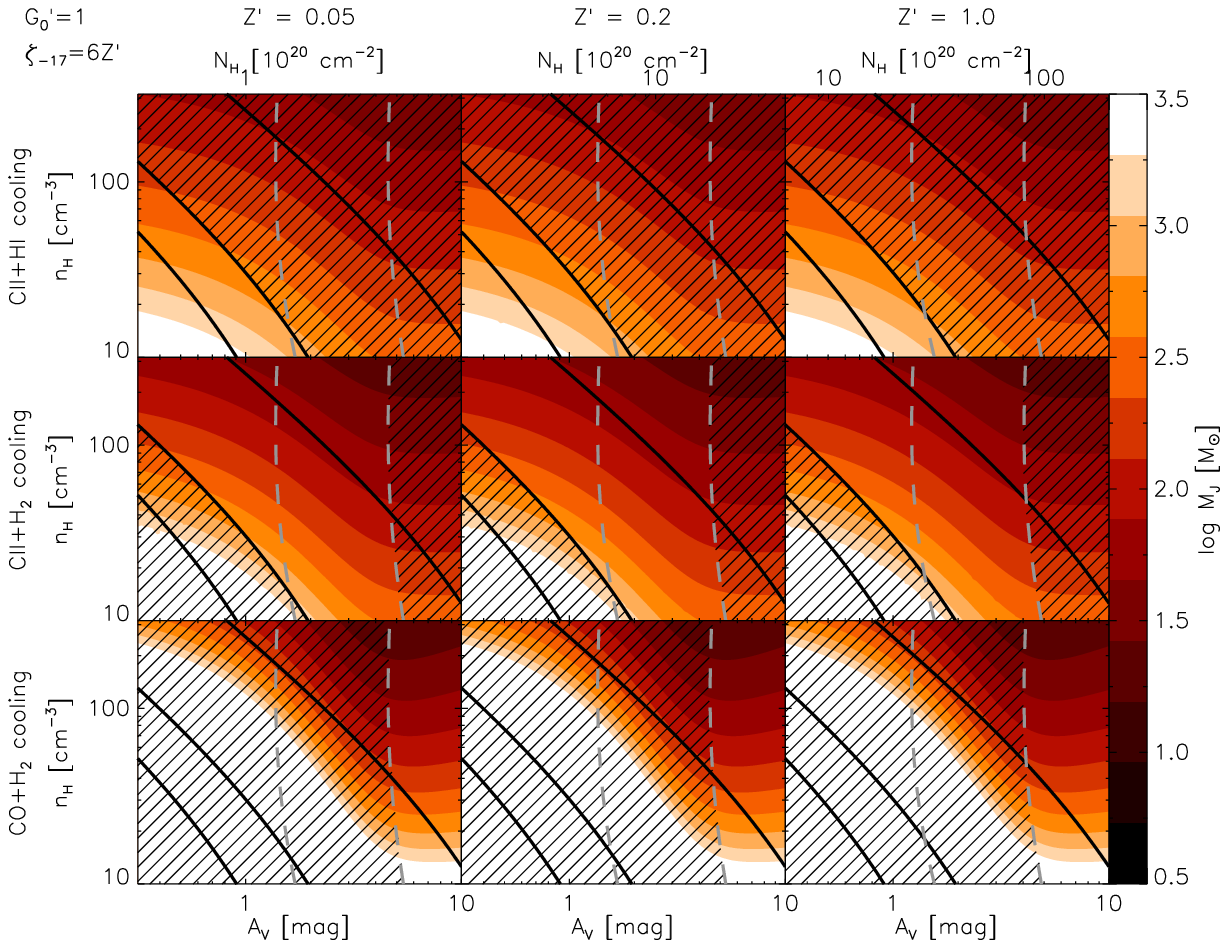}
\plottwo{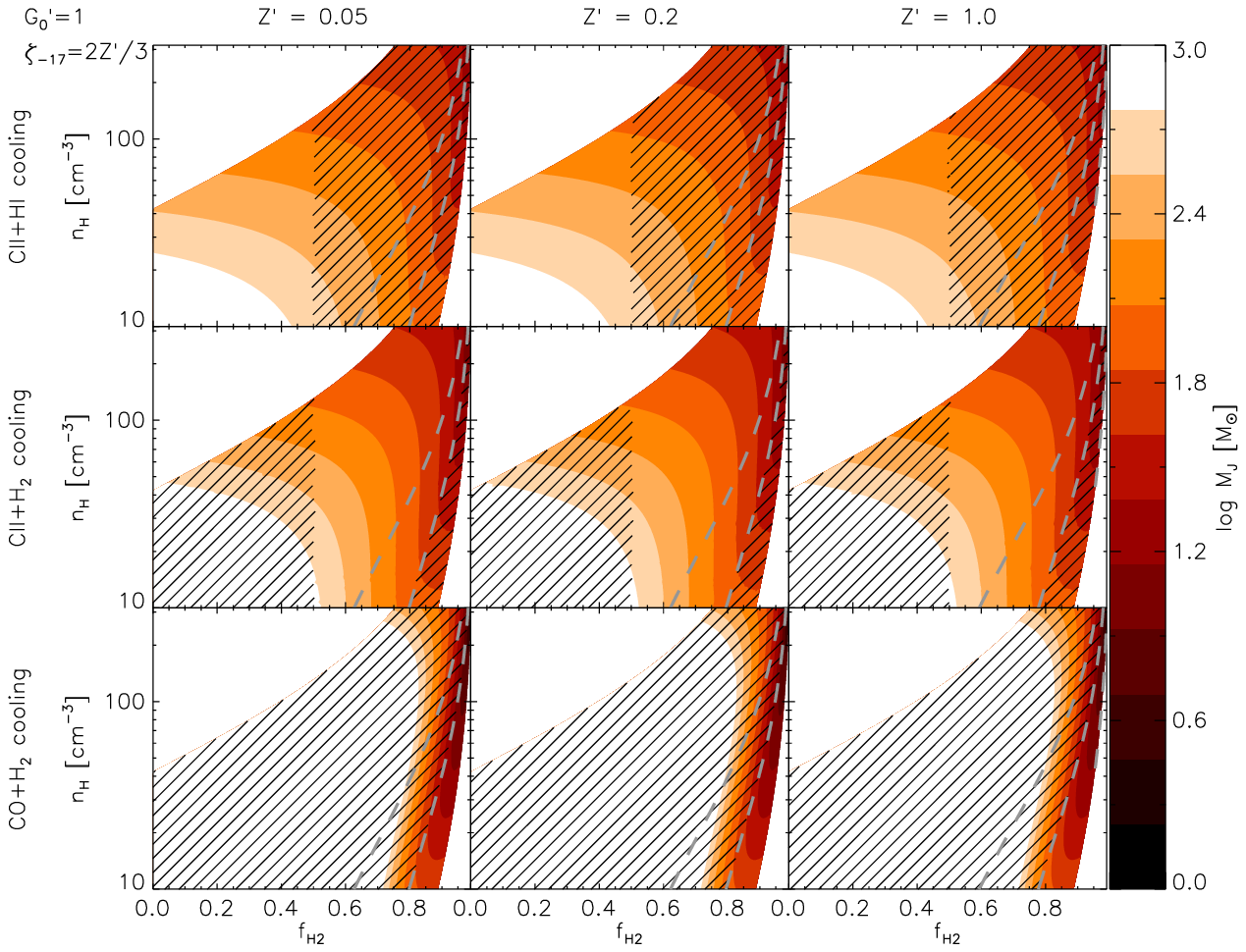}{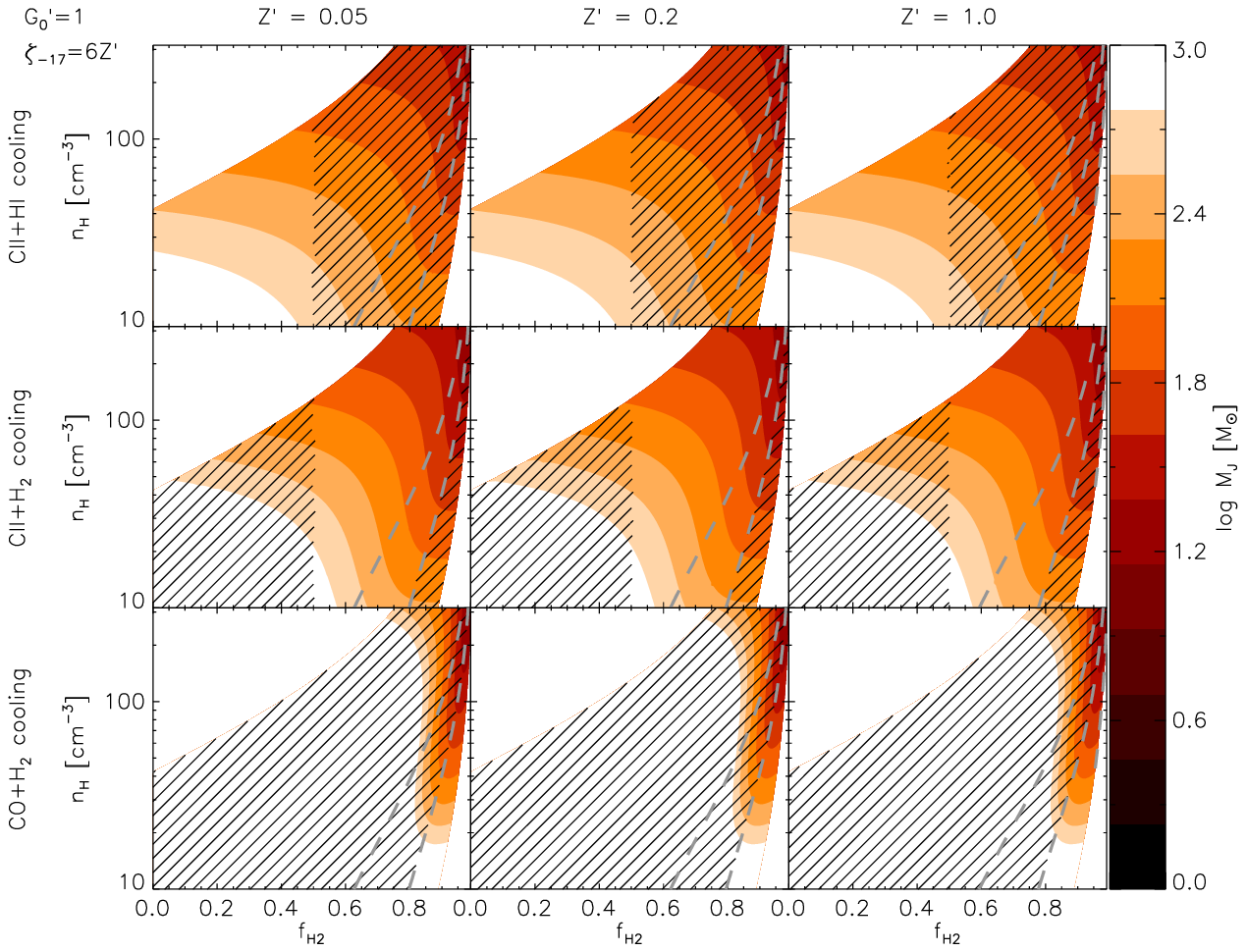}
\caption{
\label{fig:crgrid}
Left column: same as Figures \ref{fig:tempgrid}, \ref{fig:mjgrid}, and \ref{fig:mjh2}, but with $\zeta=0.67\times 10^{-17}$ s$^{-1}$ instead of $\zeta=2\times 10^{-17}$ s$^{-1}$. Right column: same as Figures \ref{fig:tempgrid}, \ref{fig:mjgrid}, and \ref{fig:mjh2}, but with $\zeta=6\times 10^{-17}$ s$^{-1}$ instead of $\zeta=2\times 10^{-17}$ s$^{-1}$
}
\end{figure}

To test our sensitivity to our choice of cosmic ray ionization rate, we have recomputed our temperature grid with cosmic ray ionization rates that are a factor of 3 larger and a factor of 3 smaller than our fiducial choice, thereby exploring a $\sim 1$ decade range in cosmic ray heating rate. We plot the results in Figure \ref{fig:crgrid}. Again, we see that the results are robust, in the sense that contours of constant Bonnor-Ebert mass remain closely aligned with contours of constant H$_2$ fraction as we vary the cosmic ray ionization rate.

We do caution, however, that this breaks down if we select a cosmic ray ionization rate that is extremely high (more than $10-30$ times our fiducial one). For such high cosmic ray ionization rates, cosmic ray heating becomes more important than grain photoelectric heating even in unshielded, low $A_V$ gas. In this case the temperature and Bonnor-Ebert mass become uncorrelated with $A_V$, and vary with density alone. However, such high cosmic ray ionization rates are inconsistent with observations in the Milky Way.

\end{appendix}

\bibliography{refs}
\bibliographystyle{apj}

\end{document}